\documentclass[USenglish,twocolumn]{article}

\usepackage[utf8]{inputenc}
\usepackage[T1]{fontenc}
\usepackage{lmodern}
\usepackage[sort&compress,square,numbers]{natbib}
\usepackage[big,online]{dgruyter}
\usepackage{xcolor}
\usepackage{bm}
\usepackage{hyperref}

\theoremstyle{dgthm}

\theoremstyle{dgdef}

\usepackage{listings}
\usepackage{xcolor}
\usepackage{amsmath}
\DeclareMathOperator*{\argmax}{arg\,max}

\definecolor{codegreen}{rgb}{0,0.6,0}
\definecolor{codegray}{rgb}{0.5,0.5,0.5}
\definecolor{codepurple}{rgb}{0.58,0,0.82}
\definecolor{backcolour}{rgb}{0.95,0.95,0.92}

\newcommand{\paramOpt}{\phi}
\newcommand{\paramGen}{\theta}
\newcommand{\paramDes}{\phi}
\newcommand{\paramFab}{\eta}
\newcommand{\paramMeasure}{\rho}

\newcommand{\x}{x} 
\newcommand{\y}{y} 
\newcommand{\de}{\chi} 
\newcommand{\m}{\upsilon} 
\newcommand{\M}{\Upsilon} 

\newcommand{\distData}{\sModelGen_\text{data}(\x)}
\newcommand{\distJointData}{\sModelGen_\text{data}(\x, \y)}
\newcommand{\distCondData}{\sModelGen_{\text{data}}(\y \vert \x)}
\newcommand{\distSimOptical}{\y(\x)}

\newcommand{\sModelGen}{p} 

\newcommand{\sModelDes}{p} 
\newcommand{\sModelFab}{r} 
\newcommand{\sModelMeas}{m} 

\newcommand{\sSpaceDes}{\mathcal{X}}

\newcommand{\modelGen}{\sModelGen_\paramGen(\x)}
\newcommand{\modelDisc}{\sModelGen_\paramGen(\y \vert x)}
\newcommand{\modelFab}{r_\paramFab(\chi \vert \x)}
\newcommand{\modelMeasure}{\sModelMeas_\paramMeasure(\m \vert \de)}

\newcommand{\fom}{f}

\newcommand{\modelDetNum}{\hat{\y}_\paramDes(\x)}
\newcommand{\modelNum}{\hat{\y}(\x)}

\lstdefinestyle{mystyle}{
  backgroundcolor=\color{backcolour}, commentstyle=\color{codegreen},
  keywordstyle=\color{magenta},
  numberstyle=\tiny\color{codegray},
  stringstyle=\color{codepurple},
  basicstyle=\small,
  breakatwhitespace=false,         
  breaklines=true,                 
  captionpos=b,                    
  keepspaces=true,                 
  numbers=left,                    
  numbersep=5pt,                  
  showspaces=false,                
  showstringspaces=false,
  showtabs=false,                  
  tabsize=2
}
\DeclareUnicodeCharacter{2212}{-}

\newcommand{\bcl}[1]{{\color{orange}{}}}
\begin{document}

\articletype{Review Article}

\author[1,2]{Yuheng Chen$^{\dagger}$}
\author[3,4]{Alexander Montes McNeil$^{\dagger}$}
\author[5]{Taehyuk Park$^{\dagger}$}
\author[1,2]{Blake A. Wilson$^{\dagger}$}
\author[1,2]{Vaishnavi Iyer}
\author[1]{Michael Bezick}
\author[1,2]{Jae-Ik Choi}
\author[1]{Rohan Ojha}
\author[1]{Pravin Mahendran}
\author[1,2]{Daksh Kumar Singh}
\author[1]{Geetika Chitturi}
\author[1,2]{Peigang Chen}
\author[1]{Trang Do}
\author[1]{Alexander V. Kildishev}
\author[1,2]{Vladimir M. Shalaev}
\author[6]{Michael Moebius}
\author[5,7]{Wenshan Cai*}
\author[3,8]{Yongmin Liu*}
\author[1,2]{Alexandra Boltasseva*}
\affil[1]{Elmore Family School of Electrical and Computer Engineering, Birck Nanotechnology Center, and Purdue Quantum Science and Engineering Institute, Purdue University, West Lafayette, IN, 47907, USA}
\affil[2]{Quantum Science Center, Oak Ridge National Laboratory, Oak Ridge, TN 37830, USA}
\affil[3]{Department of Electrical and Computer Engineering, Northeastern University, Boston, Massachusetts 02115, USA}
\affil[4]{Draper Scholar, Charles Stark Draper Laboratory, Cambridge, Massachusetts 02139, USA}
\affil[5]{School of Electrical and Computer Engineering, Georgia Institute of Technology, Atlanta, GA 30332, USA}
\affil[6]{Charles Stark Draper Laboratory, Cambridge, Massachusetts 02139, USA}
\affil[7]{School of Materials Science and Engineering, Georgia Institute of Technology, Atlanta, GA 30332, USA}
\affil[8]{Department of Mechanical and Industrial Engineering, Northeastern University, Boston, Massachusetts 02115, USA}

\title{Machine-Learning-Assisted Photonic Device Development: A Multiscale Approach from Theory to Characterization}
\runningtitle{Title}
\abstract{
Photonic device development (PDD) has achieved remarkable success in designing and implementing new devices for controlling light across various wavelengths, scales, and applications, including telecommunications, imaging, sensing, and quantum information processing.
PDD is an iterative, five-step process that consists of: i) deriving device behavior from design parameters, ii) simulating device performance, iii) finding the optimal candidate designs from simulations, iv) fabricating the optimal device, and v) measuring device performance.
Classically, all these steps involve Bayesian optimization, material science, control theory, and direct physics-driven numerical methods.
However, many of these techniques are computationally intractable, monetarily costly, or difficult to implement at scale. 
In addition, PDD suffers from large optimization landscapes, uncertainties in structural or optical characterization, and difficulties in implementing robust fabrication processes.
However, the advent of machine learning over the past decade has provided novel, data-driven strategies for tackling these challenges, including surrogate estimators for speeding up computations, generative modeling for noisy measurement modeling and data augmentation, reinforcement learning for fabrication, and active learning for experimental physical discovery.
In this review, we present a comprehensive perspective on these methods to enable machine-learning-assisted PDD (ML-PDD) for efficient design optimization with powerful generative models, fast simulation and characterization modeling under noisy measurements, and reinforcement learning for fabrication. This review will provide researchers from diverse backgrounds with valuable insights into this emerging topic, fostering interdisciplinary efforts to accelerate the development of complex photonic devices and systems.

}
\journalname{Nanophotonics}
\journalyear{2025}
\journalvolume{aop}

\runningauthor{Yuheng Chen, et al.}
\runningtitle{Machine-Learning-Assisted Photonic Device
Development}
\maketitle

\subsection*{Corresponding authors}
\noindent
\textbf{Wenshan Cai} \\
E-mail: \href{mailto:wcai@gatech.edu}{wcai@gatech.edu} \\[6pt]
\textbf{Yongmin Liu} \\
E-mail: \href{mailto:y.liu@northeastern.edu}{y.liu@northeastern.edu} \\[6pt]
\textbf{Alexandra Boltasseva} \\
E-mail: \href{mailto:aeb@purdue.edu}{aeb@purdue.edu}
\newcommand\blfootnote[1]{%
  \begingroup
  \renewcommand\thefootnote{}\footnote{#1}%
  \addtocounter{footnote}{-1}%
  \endgroup
}
\blfootnote{$\dagger$ These authors contributed equally to this work.}

\section{Introduction} 

Photonics studies light-matter interactions that encompass phenomena ranging in scale from macroscopic propagation in optical fibers to nanoscale interactions in photonic crystals, plasmonics, metamaterials, and quantum dots. The precise control of light has been made possible by engineering photonic structures, leading to the development of photonic devices for various applications, including telecommunications \citep{eldada_advances_2001, yamada_high-performance_2014}, imaging \citep{rogers_universal_2021}, sensing \citep{passaro_recent_2012,shahbaz_concise_2023}, and quantum information processing \citep{flamini_photonic_2019}. 
Photonic device development (PDD) encompasses the design, fabrication, and characterization of photonic devices to manipulate these phenomena and achieving targeted electromagnetic responses for remarkable technologies such as quantum sensors \citep{tudela24} and medicine \citep{raimondi2012two}. 

\begin{figure*}
    \centering
    \includegraphics[width=\linewidth]{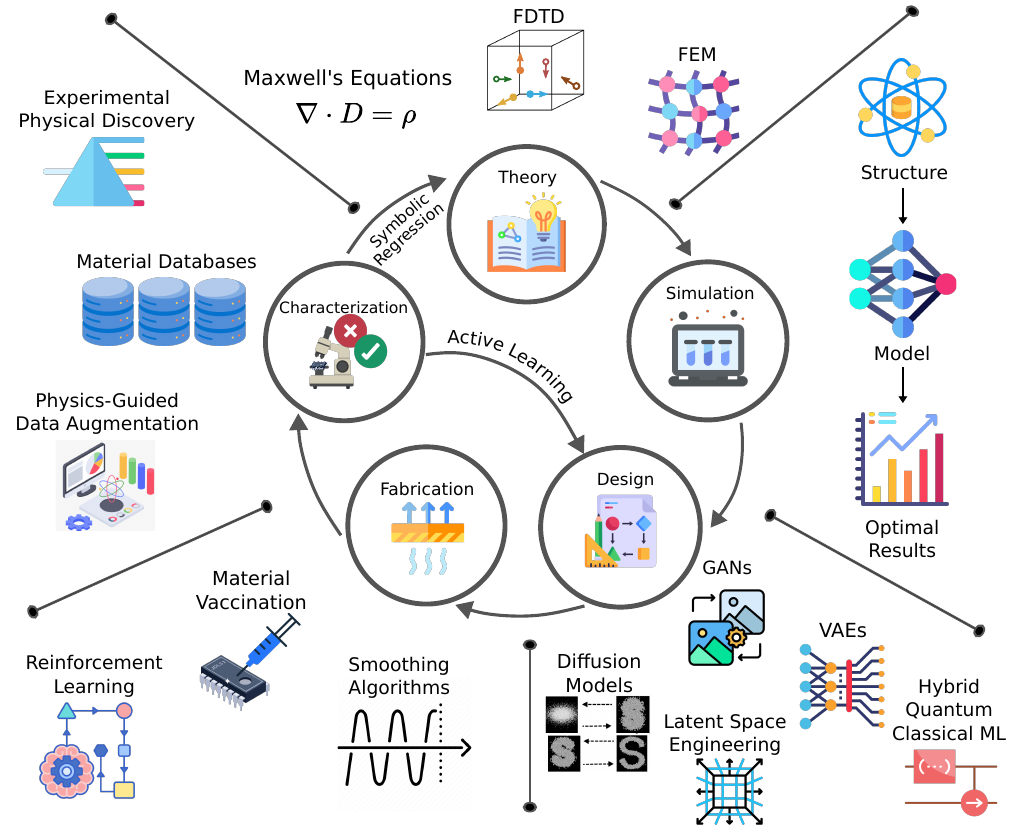}
    \caption{\textbf{Overview of Photonic Device Design (PDD) Process from Theory to Characterization.}
    This iterative process begins by deriving a system of equations that computes the optical response for each device using electromagnetic theory.
    Secondly, the design's performance and material properties are simulated either using a numerical framework, such as finite difference time domain (FDTD) or rigorous coupled-wave analysis (RCWA), or a pre-trained discriminative model.
    Then, the material distribution and its properties are generated in the design step using generative architectures such as VAEs, GANs, Diffusion models, and recently, hybrid quantum-classical models.
    Machine learning assisted inverse design techniques use simulation results to predict more optimal designs until the design is desired to be fabricated.   
    The optimal design from the generative model is then used as an ideal design for fabrication, where reinforcement learning, material vaccination, and smoothing algorithms can improve the fabrication quality. 
    Lastly, the fabricated design is characterized through several measurements. 
    High quality characterization measurements can be added to a material database, used for generative modeling and data augmentation, and utilized by active learning methods to choose better fabrication parameters.
    Completing the design cycle, physical discovery methods like hypothesis learning and symbolic regression provide surrogate models for understanding the physics of new phenomena from measurement data. 
    }
    \label{fig:main}
\end{figure*}

Although PDD is often a non-trivial task, it can be deconstructed into five major steps: i) theory, ii) simulation, iii) design, iv) fabrication and v) characterization (Figure \ref{fig:main}).
In the theory step, a model, often derived from Maxwell's equations with multiphysics coupling, is chosen to characterize the device.
These models include all the design parameters that are optimized for, including any necessary material distribution parameters, electromagnetic sources, etc.
Then, a simulation is constructed to characterize the device from its design parameters using numerical methods, such as finite-difference time-domain \citep{yee1966numerical}, finite element method \citep{clough1960finite}, and rigorous coupled wave analysis \citep{moharam1981rigorous}.
Following the simulation, the design step uses human intuition, gradient descent, or global optimization methods to propose several designs and evaluate their performances via a figure of merit (FOM) score derived from the simulation.
Designs encompass a vast parameter search space that is intractable to search exhaustively, so the design step is often computationally expensive \citep{kudyshev_global_optimization}.
As such, a significant amount of time is spent generating designs and numerically evaluating their performance until a sufficiently optimal design is obtained for the final fabrication and characterization steps.
However, fabrication processes are difficult to optimize, and designs are often finalized without fabrication tolerances in mind.
In addition, even a well-fabricated design can be subject to noisy or imprecise measurements when measuring the FOM of the fabricated device.

In the last decade, the field of machine learning (ML) has experienced unprecedented growth with new generative techniques, models, and optimizers to help photonics researchers tackle these complex problems \citep{genty_machine_2021, kudyshev_machine_2021, liu_tackling_2021, jiang_deep_2021, ma_deep_2021, xu_software_2023}.
ML techniques can learn from time-consuming simulations to propose new parameters and designs that improve fabrication quality, reduce characterization noise, and discover numerical representations of physical trends for new materials. These models enable a human designer to select between well-optimized designs and greatly reduces the volume of data and designs an experienced designer must sift through.
With the introduction of material databases, reinforcement learning, self-driving labs, hypothesis learning, and several new generative models, many of these problems are being actively addressed with significant success.

As such, this review explores ML-assisted PDD (ML-PDD), or the use of ML throughout the PDD process. 
ML-PDD contextualizes the five steps of PDD in the same Bayesian framework as modern ML, providing a structure for future work to build on.
Each section of this review is based on a step of ML-PDD and explores the application of emerging ML techniques both from recent photonics works and within the larger ML community, which may benefit future work.
Section \ref{sec:theory} explores the integration of ML into the development and implementation of EM theory, including symbolic regression to derive physical laws from experimental data, such as material property estimation, and how ML models enhance interpretability and provide physical insight to theory. Then, in Section \ref{sec:sims}, discriminative models are discussed as surrogate forward simulators, while generative models are shown to reduce the data requirements to train accurate simulation networks, thereby improving efficiency.
Next, Section \ref{sec:design} explores ML generative design strategies for effectively exploring high-dimensional design spaces, including the use of generative models such as GANs, VAEs, and diffusion models. Topics such as dimensional compression, latent space engineering, and addressing one-to-many mapping problems are also covered. Hybrid quantum-classical models are introduced as emerging tools for tackling complex design challenges.
Subsequently, Section \ref{sec:fab} addresses the challenges introduced by real-world fabrication errors absent in simulations. ML proves an ideal candidate to model the complex stochastic process that arise during fabrication, enabling more robust designs.
Following that, Section \ref{sec:characterization} views characterization as an ML inference task to deduce optical properties from experimental measurements. Challenges posed by limited training data are mitigated through physics-informed data augmentation techniques and active learning approaches, enhancing the reliability of ML models in extracting insights from sparse or noisy measurement data.
Finally, Section 7 summarizes how the integration of ML can enhance each step of PDD and offers a comprehensive outlook on the future potential of ML-PDD.

We now introduce ML-PDD and its necessary ML background information.
The authors recommend these sources as background material for a deeper understanding of the statistical framework and electromagnetic theory used in ML-PDD \cite{bishop2006pattern, vemuri2020hundred, saleh2019fundamentals, novotny2012principles}. 



   
\subsection{Background}
Photonics is deterministic in the classical limit of Maxwell's equations but designing, fabricating, and measuring real materials can be a noisy, probabilistic process because of large design spaces, fabrication errors, and noisy measurements.
Modern ML is especially powerful at learning, optimizing, and manipulating these probabilistic processes, since this is what it was originally built for \citep{ng2001discriminative, bishop2006pattern}.
ML starts by assuming that the data $\x \sim \distData$ are drawn from a data distribution $\distData$, e.g., measurement errors, material distributions, spectra, etc. 
Classically, \textit{generative models} $\modelGen$, e.g., unconditioned variational autoencoders (VAEs) \citep{kingma_auto-encoding_2022}, generative adversarial networks (GANs) \citep{goodfellow_generative_2020}, and diffusion models \citep{ho_denoising_2020}, aim to learn the joint distribution $\distData$ of all random variables in the data via maximizing the log-likelihood
\begin{align}
\argmax_\paramGen\mathbb{E}_{\x \sim \distData}[\log \modelGen], \label{eq:mle}
\end{align}
with various methods, including direct analysis, expectation maximization, and gradient descent \citep{bishop2006pattern}.
Generative models have found applications everywhere a random process needs to be modeled, e.g., simple design generators (Section \ref{sec:design}), measurement noise (Section \ref{sec:characterization}), and data augmentation (Section \ref{sec:characterization}). 

One limitation of generative models is that they are inherently restricted to learning the joint distribution of the data \citep{ng2001discriminative}.
For example, data $\x$ and their labels $\y$ have a joint distribution $\distJointData$, such as measurements and designs \cite{burnap_estimating_2016}, measurement errors and measurement parameters \cite{dhar_modeling_nodate}, fabrication artifacts and ideal designs \cite{faruqi_style2fab_2023}.
However, often a goal in ML-PDD is to train a \textit{discriminative model} $\modelDisc$ to learn the conditional distribution $\y \sim \distCondData$ for classification, regression, and conditional generative tasks such as classifying material properties \citep{yesilyurt_fabrication-conscious_2023}, characterizing structural defects \citep{zhou24}, and predicting the optical responses \citep{RN58} of generated devices. 
Discriminative modeling is widely used when only certain aspects of a variable are of interest, such as a subset of devices $\x$ or performance metrics $\y$.
For example, if a generative model $\modelGen$ generates designs in a subset, a discriminative model $\modelDisc$ can be trained to more accurately compute the labels $\y$ for designs sampled in that subset.
Generally, discriminative models perform better in practice and much of modern ML, including large language models \cite{openai2024gpt4technicalreport}, diffusion models \cite{ho_denoising_2020, rombach_high-resolution_2022}, etc. involve some form of conditional distribution \citep{ng2001discriminative}.
Another advantage of discriminative models is that they can encompass varying degrees of randomness because they are conditional. 
In other words, they are often treated as parameterized, deterministic functions $\modelDisc = \delta_y(\modelDetNum)$, where $\modelDetNum$ is a multi-layer perceptron \citep{rumelhart1986learning}, transformer \citep{vaswani_attention_2017}, or convolutional neural network \citep{krizhevsky2012imagenet}.
Discriminative models have found uses all throughout ML-PDD, including efficient forward modeling of optical systems, providing real-time predictions of device behavior without requiring full numerical simulations (Section \ref{sec:sims}), and conditional generation of designs from additional design criteria (Section \ref{sec:design}).

Many processes in PDD have distributions that belong to a known distribution family \citep{bishop2006pattern}, e.g., Gaussian noise belongs to exponential families, etc., but the parameters of the distribution are unknown, such as Gaussian noise $\mathcal{N}(\mu, \sigma)$ with unknown mean $\mu$ and variance $\sigma^2$.
For these simpler distributions, it is more effective to learn the parameters of the distribution using maximum likelihood estimator models , e.g., $\x \sim \mathcal{N}(\hat{\mu}, \hat{\sigma})$ where $\hat{\mu} = \sum_i \frac{x^{(i)}}{n}$ and $\hat{\sigma} = \sum_i \frac{(x^{(i)} - \hat{\mu})^2}{n}$, than it is to use a large machine learning model \citep{bishop2006pattern}. 
Maximum likelihood estimation is, in spirit, the driving force behind training both generative models and discriminative models. 
Each model has a set of parameters, e.g., $p_\theta$ has $\theta$, that has to be optimized to sample new data from the distribution $p_\text{data}$. 
However, for many tasks with larger, unknown distribution families, such as in inverse design (Section \ref{sec:design}), increasingly complex models are required to capture the nuances of the data.
As the complexity of the model increases, estimating the optimal parameters becomes increasingly difficult, hence the explosion of large-scale, specialized hardware and distributed techniques for training these large models \cite{openai2024gpt4technicalreport}.
Modern generative ML models and training techniques exploit large compute \cite{DBLP:journals/corr/abs-1912-01703, openai2024gpt4technicalreport} and gradient-based optimization \cite{kingma2017adammethodstochasticoptimization} to find the optimal parameters $\theta$ for optimizing Eq. \ref{eq:mle} and its many variants \citep{ng2001discriminative, bishop2006pattern}.
As such, ML-PDD takes advantage of these techniques to tackle difficult problems in PDD.

\subsection{ML-Assisted Photonic Device Development (ML-PDD) Framework}
ML-PDD begins by deriving the optical properties of a photonic device from its parameters (see Section \ref{sec:theory}). 
Each device is represented with a parameter vector $\x \in \mathcal{X}$ in a design space $\mathcal{X}$ that holds unique information about the design, for example, material topology height maps $\mathcal{X} \subseteq [0,1]^{n \times n}$ 
 \citep{wilson_machine_2021, wilson24raptor}, spectra $\mathcal{X} \subseteq \mathbb{R}^{n}$ \citep{bandi_power_2023}, voltage biases $\mathcal{X} \subseteq [0, 5]^{n}$, etc.
Each device $x$ has a corresponding optical response $y(x)$ usually derived from Maxwell's equations and any multiphysics coupling. 
Often, there is an ideal optical response $y^*$ which the design is optimized to realize, e.g., emission spectra $\y^* \in \mathbb{R}^n$ or 2D phase image $y^* \in [0, 2\pi)^{n\times n}$.
In the simulation step, as discussed in Section \ref{sec:sims}, the optical response $y(x)$ is approximated with a numerical model $\hat{y}(x)$ derived from computational electromagnetic techniques \citep{bondeson2012computational}.
A deterministic model $\hat{y}_\phi(x)$ with parameters $\phi$ known as a \textit{surrogate estimator model} (Section \ref{sec:sims}) is trained to approximate the numerical model $\hat{y}(x)$, speeding up the inference time by orders of magnitude at the cost of some accuracy.
The surrogate estimator model is deterministic but it can be augmented to model noisy measurements, for example Gaussian noisy measurements $\y = \modelDetNum + \epsilon : \epsilon \sim \mathcal{N}(\mu, \sigma)$.
Taking the surrogate estimator model $\hat{y}_\phi(x)$ and the ideal optical response $y^*$, the following design step (Section \ref{sec:design}) constructs a figure of merit (FOM) score function $f(y)$ to rank designs based on their optical response, e.g., $f(y) = \vert\vert y^* - y\vert\vert_l$ for some $l$-norm.
Then, a generative model $x \sim p_\theta(x)$ is trained, see Section \ref{sec:design}, to sample designs that maximize the FOM in the expectation 
\begin{align}
    \argmax_\theta \mathbb{E}_{x \sim p_\theta(x)}[f(\hat{y}_\phi(x)) ]\label{eq:des_opt_fom}.
\end{align}
Often, the design step requires several iterations $t = 1,...,T$ of sampling designs $x^{(t)} \sim p_{\theta^{(t)}}(x)$, evaluating their performance $f(\hat{y}_\phi(x^{(t)}))$, and retraining the generative model ${\theta^{(t)}} \rightarrow {\theta^{(t+1)}}$ before choosing a sufficiently optimal set of designs $X^* = \{x^{(t_i)}\}_{i=1}^N$ for fabrication.
Given these optimized designs $X^*$, a probabilistic fabrication process $r_\eta(\chi \vert x)$, see Section \ref{sec:fab}, with fabrication parameters $\eta$, e.g., alignment, tolerances, etc., takes an intended design $\x \in X^*$ and fabricates a device $\chi$.
Fabrication is a very difficult and time-consuming process that cannot fully realize a design $\x$. 
Therefore, techniques such as reinforcement learning have been crucial in reducing randomness and improving the final quality of designs, which will be discussed in Section \ref{sec:fab}.
Once the design is fabricated $\de \sim \modelFab$, the optical response of the fabricated design is measured via a noisy measurement device $\m \sim \modelMeasure$ with noisy measurements $\m$ and measurement parameters $\paramMeasure$, as discussed in Section \ref{sec:characterization}.
Naturally, to accommodate noisy measurements, the FOM is augmented $\hat{f}$ to use a finite sample of noisy measurements $\M = \{\m^{(i)} \sim \modelMeasure \}_{i=1}^M$.
The overall objective of ML-PDD is to optimize the device performance in this noisy environment
\begin{align}
    \argmax_{\paramGen, \paramFab, \paramMeasure} \mathbb{E}_{\x \sim \modelGen}[\mathbb{E}_{\de \sim \modelFab}[\mathbb{E}_{\m \sim \modelMeasure}[\hat{f}(\M)]]] \label{eq:pdd_full}.
\end{align}
To make the ML-PDD objective tractable, assumptions are made on the fidelity of measurements, the optimality of designs and fabrication processes.
Steps are isolated and optimized. For example, in inverse design, often the numerical simulation FOM is treated as the ground-truth without considering fabrication processes $\modelFab$ or noisy measurements $\modelMeasure$.
As such, the remainder of this review will explore each step both in isolation and coupled to other steps, broken down into theory (Section \ref{sec:theory}), simulation (Section \ref{sec:sims}), design (Section \ref{sec:design}), fabrication (Section \ref{sec:fab}) and characterization (Section \ref{sec:characterization}).

%
%
%
%

\section{Theory} 
\label{sec:theory} 

Classically, electromagnetic (EM) theory has been developed through experimental observations, theoretical reasoning, and mathematical synthesis
 \citep{yu_plasmon-enhanced_2019,shastri_photonics_2021}.
Nowadays, applying EM theory to investigate specific systems often involves large computational requirements while struggling to fully capture the richness of photonic phenomena \citep{shastri_photonics_2021}.
ML offers a compelling alternative by enabling the discovery of hidden mathematical relationships \citep{shu_knowledge_2023}, predicting material properties with remarkable accuracy \citep{liu_materials_2017}, and solving intricate EM systems that defy conventional techniques \citep{li_empowering_2022}, empowering researchers to rethink the discovery of governing physical laws and enhance the interpretability of results.
This section explores developments for the derivation of optical response $\distSimOptical$, emphasizing the new role that ML may play in the development of future photonic EM theories. 
Specifically, in discovering new governing laws and interpretable ML for characterization $\distSimOptical$ via symbolic regression and explainable AI. 

%


\subsection{Discovering Governing Laws} 

A central application of ML in EM theory is the discovery of governing equations and physical laws to model optical responses.
Traditional approaches to understanding EM phenomena often involve laborious analytical derivations or heuristic methods informed by experimental observations \citep{shastri_photonics_2021}.
ML models, particularly symbolic regression and neural networks, automate this process by identifying mathematical relationships within data \citep{li_empowering_2022}.
A rapidly growing Python implementation known as PySR \cite{cranmer2023interpretablemachinelearningscience} combines symbolic regression with genetic programming to construct analytic expressions. Similar techniques have successfully rediscovered Maxwell’s equations by analyzing datasets of electric and magnetic field values in various scenarios \citep{kim_integration_2021,alnuqaydan2023symbolic}.
Beyond rediscovering well-established laws, ML has been employed to hypothesize new governing equations in complex environments, such as through experimental physical discovery. 
For instance, ML has identified novel constitutive relationships that better describe the interaction of fields and materials in metamaterials and plasmas, where traditional models often struggle due to high degrees of complexity and anisotropy \citep{li_deep_2022}. These breakthroughs enable more accurate predictions and a deeper understanding of exotic materials and devices.
For example, neural networks trained on EM field distributions and boundary conditions can infer constitutive relations for materials \citep{tartakovsky_learning_2018,jiang_deep_2020}, such as the relationship between electric field intensity and polarization in nonlinear media of photonic devices \citep{genty_machine_2021}.

\subsection{Enhancing Interpretability and Physical Insight}
\begin{figure*}
    \centering
    \includegraphics[width=\linewidth]{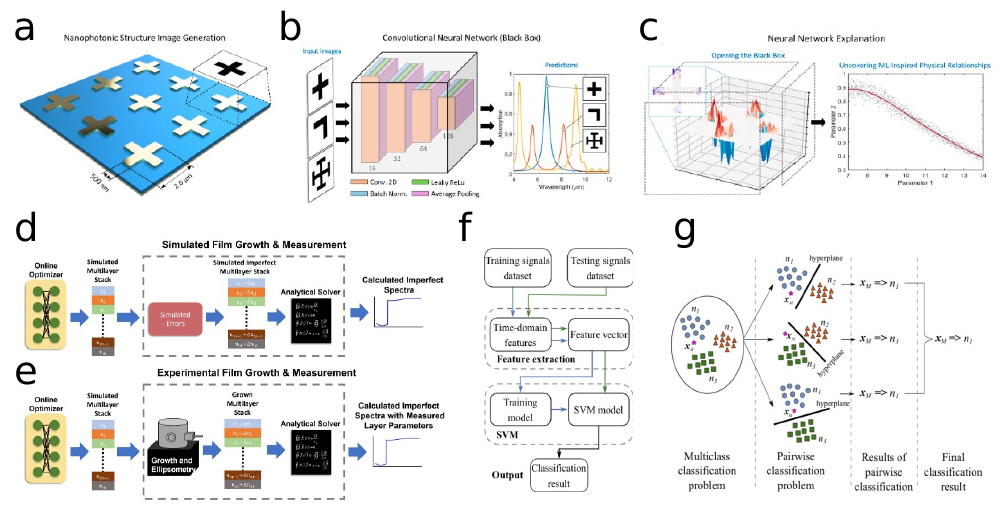}
    \caption{\textbf {Machine learning applications in photonic theory.} Three steps of explainable ML elucidating the behavior of nanophotonic structures:  a) Converting 3D metal-dielectric-metal metamaterials into 2D representations. b) Training a convolutional neural network (CNN) to predict the electromagnetic response. c) Elucidating the underlying physics learned by explaining the relationships between structural features and predicted parameters to construct new designs with new target properties. Subfigures a, b, and c are adapted with permission from \cite{yeung_elucidating_2020}. Copyright 2020 American Chemical Society. d) Network architecture for fabrication-in-loop NN-based inverse design of single-material multilayer optical stacks with continuously changing refractive index in simulated fabrication case. e) Measured and experimentally retrieved layer parameters are integrated back into the optimization cycle. Subfigures d and e are adapted with permission from \cite{yesilyurt_fabrication-conscious_2023}. Copyright 2023 the author(s), published by De Gruyter, Berlin/Boston, licensed under the Creative Commons Attribution 4.0 International License. f) Flowchart of the SVM model and g) diagram of the classification process in unambiguous refractive index measurement. Subfigures f and g adapted with permission from \cite{martinez-manuel_machine_2022}, Copyright 2022, IEEE.
    }
    \label{fig:theory}
\end{figure*}

One of the key challenges in ML-PDD, in general, is integrating EM theory into training to ensure that models are interpretable and aligned with physical principles \citep{yeung_elucidating_2020,elzouka_interpretable_2020}.
Advances in ML techniques, such as attention mechanisms and feature importance analysis \citep{shao_machine_2024}, have made it possible to reveal explanations for models' decisions and enhance our understanding of the underlying physics.
For example, Yeung et al., in Figure \ref{fig:theory} (a)-(c), used a three-step explainable ML approach to uncover the relationship of specific regions of a nanophotonic structure to the presence of an absorption peak, leading to a better understanding of the behavior of complex nanophotonic devices and enabling more efficient and targeted design improvements \citep{yeung_elucidating_2020}.

Furthermore, by highlighting the most critical parameters, ML models could also provide actionable insights that guide iterative design processes, estimating important parameters such as permittivity, permeability, or conductivity by analyzing data from spectroscopy, reflectometry, or scattering experiments \citep{armghan_graphene_2023,yun_deep_2022}.  For instance, Yesilyurt et al. employed a discriminative ML model to extract material refractive indices and loss coefficients from the spectral transmission and reflection data for the realistic design and fabrication of single material variable-index multilayer films in Figure \ref{fig:theory} (d) and (e) \citep{yesilyurt_fabrication-conscious_2023}. The classification of modes and band structures is also essential for understanding and optimizing photonic device performance. ML models can efficiently identify guided modes, radiative modes, or photonic band gaps based on structural and material inputs to enhance our understanding of photonics devices \citep{venketeswaran_recent_2022}. For example, Martinez-Manuel et al. applied support vector machines (SVMs) (Figure \ref{fig:theory} (f)) for unambiguous refractive index measurement of the fiber fundamental mode in Figure \ref{fig:theory} (g) \citep{martinez-manuel_machine_2022}. Similarly, Li et al. utilized deep learning approaches to analyze photonic band structures of phononic crystals, identifying band gaps and key dispersion features \citep{li_designing_2020}. 

Latent space engineering in generative models, such as autoencoders, offers another avenue for interpretability (see Section \ref{sec:design}).
These models encode high-dimensional data, like photonic design geometries or spectral responses, into compact representations that reveal patterns and relationships.
For instance, clustering designs in latent space based on efficiency metrics can help researchers identify common features of high-performance devices \citep{singh_deep-learning_2024}.

\subsection{Summary and Outlook}
ML is transforming how we apply EM theory and materials science in PDD. As ML continues to evolve, its integration with foundational physics principles offers a promising avenue for breakthroughs in device performance, material characterization, and the exploration of new photonics device frontiers. However, achieving this vision requires addressing key research gaps, such as overcoming data scarcity through efficient learning methods, enhancing the robustness, interpretability, and real-time adaptability of ML models. By tackling these challenges, researchers can foster a synergistic ML-EM paradigm that drives innovation in photonics and unlocks unprecedented opportunities for scientific and technological advancement.



\section{Simulation} 
\label{sec:sims}
Optical simulations computationally approximate the optical response $y(x)$ of each design $x$ via a numerical model $\hat{y}(x)$, including reflection, transmission, absorption, scattering, EM field distribution, chirality, and polarization. 
These simulations play a crucial role in studying light-matter interactions in photonic structures and in the design and optimization of photonic devices.
In PDD, optical simulations predict the performance of designed structures and evaluate candidate structures during the iterative design process outlined in Section \ref{sec:design}.
Traditionally, the optical response of a structure has been obtained by numerically solving Maxwell's equations.
Since numerical simulations directly solve the governing equations, they are often treated as reliable and their accuracy can be improved by refining material meshes, increasing the number of harmonics in the simulation, etc. \cite{bondeson2012computational}.
However, obtaining accurate numerical solutions is computationally intensive, particularly for complicated structures and systems, and restricts the rate of design evaluation throughout the PDD.

In contrast, ML models can act as efficient and reliable surrogate forward simulators, providing a faster alternative to time-consuming numerical simulations.
ML models treat the simulation process as a discriminative regression task, mapping input structures $x$ to output characteristics $y$ based on labeled datasets of input-output pairs.
Once well-trained, these simulation ML models offer much faster computation speeds at the cost of some accuracy. 

This section discusses how ML techniques provide a fast and reliable alternative to traditional full-wave calculations.
The role of ML models in this section is categorized into two main aspects: first, how discriminative models enable computationally efficient predictions of device responses, i.e., training ML models as the approximator $\hat{y}_\phi(x)$, and second, how generative models reduce the data demands required for training ML simulator models, i.e., how to increase the quality and quantity of training data for $\hat{y}_\phi(x)$. 
These techniques demonstrate that ML-enabled, data-driven simulations are powerful and agile tools that accelerate the evaluation of complex photonic designs, making the design process more efficient.


\subsection{Surrogate Forward Simulator Models}
\begin{figure*}
    \centering
    \includegraphics[width=\linewidth]{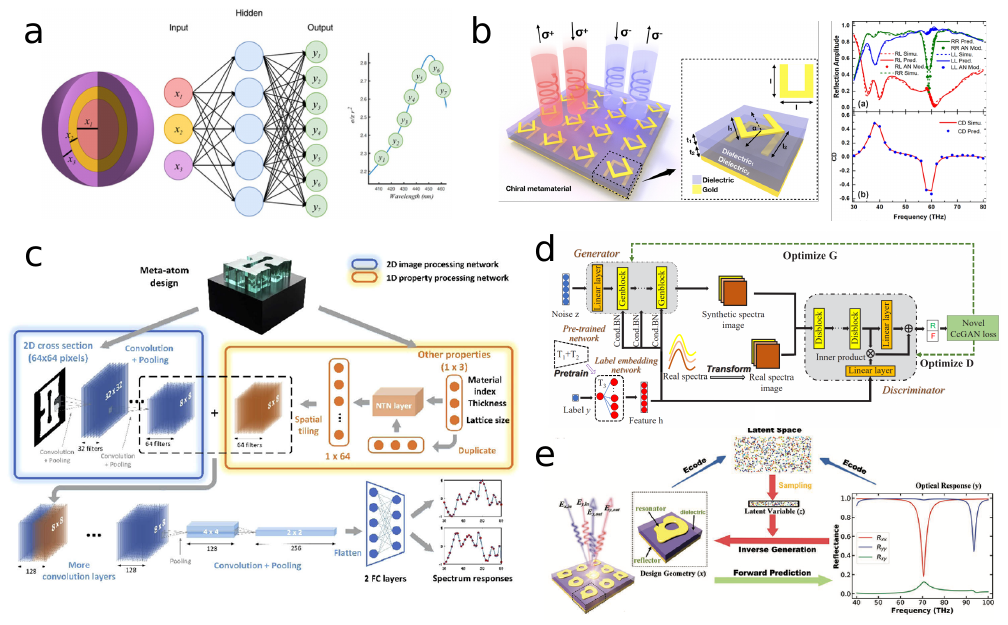}
    \caption{\textbf {Optical Simulations Using Deep Learning Methods.} a) A fully connected neural network predicts the scattering spectra of multilayered core-shell nanoparticles. Adapted with permission from ref  \cite{RN58}, Copyright 2018 The Authors, distributed under the Creative Commons Attribution-Non Commercial license.
    b) A simple geometry of chiral metamaterials, consisting of two gold twisted split-ring resonators, is represented using a few geometric parameters. A neural network model predicts circular dichroism from the given geometry. Reprinted with permission from  \cite{RN15}. Copyright 2018 American Chemical Society.
    c) Freeform meta-atoms are represented as 64 × 64 binary images. The prediction of the real and imaginary components of the spectral responses is obtained through CNN. Adapted with permission from  \cite{RN21}, Copyright 2020 Optical Society of America.
    d) The architecture of the CcGAN, which generates synthetic spectra for data augmentation. The generated spectra were conditioned on the temperature label \(y\). The performance of the ML temperature prediction model was significantly enhanced through data augmentation using the CcGAN. Adapted with permission from  \cite{RN405}, Copyright 2023 Optica Publishing Group.
    e) Self-supervised learning is applied in a VAE architecture. The encoder predicts reflection spectra as well as compresses the input structure into a latent vector, enabling the model to train with both labeled and unlabeled data. Adapted with permission from  \cite{ma_probabilistic_representation}, Copyright 2019 WILEY-VCH Verlag GmbH \& Co. KGaA, Weinheim.
    }
    \label{fig:sims}
\end{figure*}

ML-based surrogate estimators are discriminative and data-driven, involving an inference task that approximates the optical response \(\hat{y}(x)\), of a device design, \(x\).
Therefore, discriminative models that learn the complex relationship between input and output based on a training dataset of labeled input-output pairs are well-suited for simulation problems.
Since the models generate data-driven predictions of the output response from the input structure using ML, they offer orders-of-magnitude speed-up over numerical simulations, which directly solve simulation problems. In the context of ML applications for simulations, both the input and output are typically represented as vectors. For optical devices with a low degree of freedom, such as multilayer films and periodic structures containing simple geometric patterns, the structures can be fully represented by a vector \(x\) comprising a series of thicknesses or geometric parameters, as depicted in Figure \ref{fig:sims} (a) and (b). For high-degree-of-freedom structures, such as those containing freeform patterns, the structures can be represented using binary images, where materials are distinguished by ‘0’ and ‘1’, as shown in Figure \ref{fig:sims} (c) and (e). For the output, spectra, for example, can be sampled at a sufficiently fine rate to construct a vector representation of \(\hat{y}(x)\). A variety of simulation problems in optics have been demonstrated using discriminative neural networks.

The simulation of optical devices with a low degree of freedom, including layered structures \cite{RN58, liu2018training, RN75} and patterns with primitive geometries \cite{RN67, RN15}, can be conducted using the fundamental discriminative model, fully connected networks. Peurifoy et al. employed them to approximate light scattering from a dielectric spherical nanoparticle with alternating silica and titanium oxide shells \cite{RN58}. The model achieved high precision in predicting scattering spectra and generalized well to the samples that were not seen during the training, enabling it to serve as a surrogate model for forward simulation of such structures. These forward simulation networks play a crucial role in inverse design by being integrated with inverse design networks that map the optical responses to design parameters. For layered structures, these networks have facilitated the inverse design of core-shell nanoparticles to achieve specific electric and magnetic dipole extinction spectra \cite{RN75}, as well as thin-film structures of alternating SiO\(_2\) and Si\(_3\)N\(_4\) for target transmission spectra \cite{liu2018training}. Similarly, meta-atoms with primitive geometries, represented by vectors of a few geometric parameters, can also be simulated and designed using neural networks. For instance, deep learning-based simulations and inverse designs have been demonstrated for chiral metamaterials composed of two twisted gold split-ring resonators separated by dielectric spacers \cite{RN15}, as well as metasurfaces with H-shaped gold nanostructure on top of ITO-covered glass \cite{RN67}.

For the simulation of optical structures with a high degree of freedom, more advanced models are required to ensure accurate predictions. Convolutional Neural Networks (CNNs), which utilize convolution operations with kernels to extract local features and capture spatial hierarchies in input data, are well-suited for the simulation task of optical devices with complex geometries. Freeform patterns, represented as images, benefit from CNNs due to their ability to efficiently capture local correlations within the image, making them a desirable choice for such simulations. An et al. simulated the spectral responses of dielectric metasurfaces, where quasi-freeform meta-atoms were represented by 64×64 pixel images using a CNN \cite{RN21}. The trained CNN simulator demonstrated high precision in predicting the responses of not only quasi-freeform patterns in the test dataset but also generalized well to circle- and ring-shaped meta-atoms, which were entirely different geometries from those in the training dataset. Wiecha et al. successfully simulated the internal fields of arbitrary 3D nanostructures using a CNN in a U-net-like architecture with residual connections \cite{RN20}. This sophisticated model enabled accurate simulations that reproduced complex physical effects in both plasmonic and dielectric nanostructures. Furthermore, the integrated architecture of ResNet-CNN with residual connections—combined with a recurrent neural network (RNN) demonstrated a good match with numerical simulations of the absorption spectra of periodic structures consisting of silver on top of a glass substrate \cite{sajedian2019optimisation}. In this setup, the CNN extracted spatial features, while the RNN predicted the spectra based on the output from the preceding CNN.

Recently, instead of data-driven simulations that learn functions \(x \mapsto y\), which map input and output vectors, neural operators, which learn operators \(x(\boldsymbol{r}) \mapsto y(\boldsymbol{r})\) that map between functions, have gained significant attention. Neural operator architectures consist of multiple layers of linear integral operators followed by nonlinear activations \cite{RN412}. Neural operators have been applied to learn solution operators for partial differential equations (PDEs), including Darcy flow, Burgers’ equation, and the Navier–Stokes equations, enabling significantly faster simulation than conventional numerical solvers \cite{RN412, RN411}. Moreover, neural operator models generalize well across different levels of discretization, as they learn mappings between continuous functions, whereas traditional vector-to-vector neural networks are heavily dependent on the discretization scheme \cite{RN413}. The idea of using neural operators to solve PDEs has also been extended to Maxwell’s equations for electromagnetic simulations. Gu et al. proposed the NeurOLight framework, which combines a PDE encoder with a neural-operator-based backbone to build a surrogate forward solver for simulating multi-mode interference photonic devices \cite{RN416}. The NeurOLight enables efficient simulation of a family of parametric photonic devices, rather than solving only a single instance or one conditioned on fixed parameters.

\subsection{Data Augmentation using Generative Models}
While the forward simulation problem is inherently discriminative from a machine learning perspective, generative models can play a role in enhancing the data efficiency of simulator network training. Since the training of simulator networks is data-intensive, achieving highly accurate simulator networks requires a large amount of labeled data—input structure and output optical response pairs—obtained through computationally expensive numerical simulations. Reducing dependency on data can be achieved by incorporating known optical knowledge \cite{RN179, RN105, RN39} and employing machine learning techniques such as transfer learning \cite{RN52, RN184, RN96} and generative models \cite{kim_semi-supervised_2024, RN405, RN4, RN68}. Physical knowledge serves as a form of regularization during the training of simulator networks, helping the network find valid solutions with less data. Transfer learning is a technique where a model trained on a source task with a fairly large dataset is used to initialize the trainable parameters for a target task. This approach enables faster convergence and improved performance on the target task, even when the target dataset is relatively small, by leveraging the knowledge learned from the source task. Furthermore, generative models alleviate the burden of data collection by augmenting datasets through their generative capabilities and enabling self-supervised learning within encoder-decoder architectures.

Generative models enable efficient data-driven simulation by generating synthetic pseudo-labeled samples, which support semi-supervised learning for simulator networks by leveraging the generated data. Kim et al. employed a denoising diffusion probabilistic model (DDPM) to enhance the performance of a simulator network designed to predict the transmission spectra of photonic crystal waveguide unit cells \cite{kim_semi-supervised_2024}. The DDPM was trained on images of photonic crystal patterns, allowing it to generate unlabeled new patterns later labeled by a pretrained CNN simulator network. Since these newly generated patterns from the DDPM were produced by learning the probabilistic distribution of the training data, the pseudo-labeling process achieved a high-fidelity approximation using the pretrained simulator. Additionally, Zhu et al. introduced a continuous conditional GAN (CcGAN) to generate synthetic spectra of a self-interference microring resonator sensor at different operating temperatures \cite{RN405}. As illustrated in Figure \ref{fig:sims} (d), the CcGAN was conditioned on the temperature, and the corresponding generated spectra were combined with the original small training dataset for data augmentation. By leveraging the augmented data with synthetic pseudo-labeled samples for training ML forward prediction networks, the prediction accuracy improved, and the error distribution shifted toward lower losses.

Another approach to achieving efficient data-driven optical simulation using generative models is the application of self-supervised learning in encoder-decoder architectures \cite{RN4, RN68}. Ma et al. utilized a Variational Autoencoder (VAE), where the encoder not only compresses the input nanophotonic structure into a latent vector but also predicts the corresponding spectral response. This setup enables the self-supervised learning strategy to improve the performance of nanophotonic structure characterization \cite{RN4}. As illustrated in Figure \ref{fig:sims} (e), in this VAE architecture, an additional loss term for prediction accuracy was introduced, along with the reconstruction error and KL divergence. During training, both structures from readily available labeled data obtained from numerical simulations and dynamically generated structures were fed online into the encoder. The ground truth spectra were used to calculate the loss for labeled samples, while the generated structures were pseudo-labeled by the encoder itself. This approach allowed the training process to be self-supervised by the encoder. The total loss, which included the prediction error, was backpropagated through the entire encoder-decoder structure, leading to a further reduction in prediction loss compared to a fully supervised counterpart, thereby improving the simulation capability of the model.

\subsection{Physics-Inspired Constraints in Model Training}
A key advancement in ML modeling for photonics is the incorporation of physics theory-based constraints into model training. For instance, by embedding Maxwell's equations as regularization terms or constraints, ML models can ensure that their outputs adhere to fundamental physical laws  \citep{leon_physics-constrained_2024}. It not only reduces the reliance on large datasets, but also simultaneously improves the interpretability and reliability of the ML models for both simulation and design of photonic devices. Especially, surrogate simulator models $\modelDetNum$ are trained by minimizing a loss function dependent on the parameters $\paramOpt$. By incorporating physical constraints into the loss function, the surrogate estimator $\modelDetNum$ is more interpretable and physics informed, thereby better able to approximate the optical response $\modelNum$.

Physical constraints play a pivotal role in reducing the effort required for data collection in training models. In PDD, assuming symmetry, e.g., rotational or axis mirror, simplifies the design space and computational requirements. For instance, simulations can be restricted to a quarter or half of the unit cell by applying symmetric boundary conditions along the axes, depending on the type of symmetry involved. Symmetry also ensures continuity and connectivity at the boundaries between neighboring unit cells, which is critical in the design of gratings. Additionally, symmetry is often leveraged to augment training datasets and regularize machine learning models, ensuring improved performance and data efficiency. For example, in periodic structures based on optical principles, invariances in the spectrum under transformations such as translation, flipping, and 180° rotation can be exploited  \citep{RN362, RN363}. Furthermore, 90° or 270° rotations of the pattern induce cross-polarization, effectively swapping the $\x$ and $\y$ polarization components while maintaining the overall spectral properties. These symmetries enable data augmentation for simulator networks that predict optical characteristics by including structure-characterisitics pairs induced by physics constraints\citep{RN317, RN319}. This approach is not limited to discriminative simulator models but also applies to generative models for design, significantly reducing the reliance on computationally expensive numerical simulations or topology optimizations \citep{RN362, RN363}. 

Incorporating physical knowledge about resonances  \citep{RN105, RN179} and electrical field distributions  \citep{RN177, RN39} as constraints into the model training process serves as a form of regularization, further enhancing data efficiency. In optics, integrating Maxwell’s equations as a loss function has proven particularly effective for solving inverse problems. Applications include tasks such as permittivity retrieval  \citep{RN66, RN98}, designing invisible cloaking devices  \citep{RN66}, and optimizing meta-lenses  \citep{RN48, RN93}, all achieved with significantly reduced data requirements. These methods highlight the importance of embedding physical principles to improve the robustness and efficiency of machine learning models in photonics.

In addition, fabricability imposes significant physical constraints, requiring the elimination of intricate design features that are infeasible to fabricate and ensuring robustness against fabrication errors  \citep{RN362, RN363, RN333, RN54}, as we will detailedly discuss in Section \ref{sec:fab}.

\subsection{Summary and Outlook}
In this section, we explored how ML simulator models can perform simulations much faster than numerical approaches by predicting the optical response of photonic devices in a data-driven manner, rather than directly solving equations.
However, these models require training on a large amount of labeled data, which must inevitably be obtained from time-consuming numerical simulations to achieve sufficiently accurate results. Generative ML models mitigate the burden of data collection through self-supervised learning, leveraging generative techniques to synthesize labeled data, while physical constraints contribute through data augmentation and regularization during model training.  Therefore, well-trained simulator models, combined with effective data generation strategies, can expedite the iterative performance evaluation of intermediate designs, enabling computationally efficient PDD. With fast performance evaluation of devices in place, how to effectively explore the large design space and identify optimal designs will be discussed in detail in Section 4.

\section{Design} 
\label{sec:design}
Generating designs $\x \sim \modelGen$ in PDD requires sampling complicated, sparse and high-dimensional design spaces $\sSpaceDes$ to achieve specific optical responses $\y^* = \modelNum$ such as spectra, bandwidth or polarization.
Many of the classical techniques, such as adjoint optimization, physics-inspired optimization, or even evolutionary algorithms, are iterative and directly optimize Eq. \ref{eq:des_opt_fom} by first randomly proposing a design $\x^{(0)} \sim \sModelDes(\x^{(0)})$, and then iteratively proposing new designs $\sModelDes(\x^{(t+1)} \vert \x^{(t)}, t)$ until the FOM $\fom(\hat{\y}^{(t)})$, where $\hat{\y}^{(t)} = \hat{\y}_\paramDes(\x^{(t)})$, saturates at a locally optimal value.
Many classical design methods use gradient based optimizers such as adjoint optimization \cite{yeung_enhancing_2022, hughes_adjoint_2018} and gradient descent \cite{deng_deep_2022, liu_tackling_2021}, etc. and evaluate the FOM using the computationally expensive numerical model $\modelNum$.
As such, classical methods are computationally expensive, slow, and get stuck in local minima where the change in FOM $\Delta \fom(\y^{(t)}) / \Delta t \rightarrow 0$ between design proposals decreases rapidly with $t$.
On top of that, the design generation process requires a delicate balance of performance optimization and adherence to constraints, e.g., physical, geometric, and fabrication.

To address this delicate balance, generative models, such as Generative Adversarial Networks (GANs), Variational Autoencoders (VAEs), diffusion models, RL, and hybrid quantum-classical models have rapidly proven especially effective to tackle these problems.
These models are uniquely suited for exploring vast and sensitive design spaces, enabling the creation of innovative and highly customized device configurations.
They can rapidly generate solutions and learn complex relationships between design parameters and performance metrics such as size, material properties, and operational robustness, allowing for the automated discovery of non-intuitive solutions.
Additionally, generative models allow for the probabilistic modeling of design spaces, making it possible to predict and mitigate variations or uncertainties in performance. By accelerating the exploration of design possibilities and enhancing the precision of results, ML technologies, especially generative models, are reshaping the photonics design process, driving innovation, and enabling the creation of cutting-edge optical devices.
In particular, the one-to-many mapping problem is significant in photonics design because a single desired optical response can often correspond to multiple distinct designs, making traditional optimization approaches inefficient and prone to converging on suboptimal solutions. Probabilistic modeling facilitated by generative models enables the exploration of diverse, non-intuitive solutions that satisfy the same performance criteria.
ML-driven optimization techniques are growing in popularity for their ability to reduce the dimensionality of design space \cite{cai_topological_encoding, wilson_empowering_2022}, facilitating the exploration of new designs \cite{ma_probabilistic_representation, wu_meta_atom, chen_advancing_2024}, performing global optimization of photonic devices \cite{kudyshev_machine-learning-assisted_2020, kudyshev_global_optimization, mascaretti_designing_2023}, address the one-to-many mapping problem in inverse design \cite{kumar_multi_solution, li_bound_states}, and explore new physics in highly complex photonic structures \cite{wang_mechanistic_modeling}.
To prepare the use of each of these models, the FOM should be considered carefully.

\subsection{Constructing the Figure of Merit (FOM)}
\label{sec:design_fom}
A figure of merit (FOM) $\fom(\y) \in \mathbb{R}$ is a scalar measure of device performance which scores high-dimensional optical response labels $\y$ into a single metric to facilitate optimal device design. A higher FOM value for an optical response $\y_i$ compared to another response $\y_j$ ($\fom(\y_i) > \fom(\y_j)$), indicates that $\y_i$ is ``closer'' to the optimal response $\y^*$ \footnote{Or vice versa if PDD is applied to a minimization problem.}.

A common approach to defining FOMs in deterministic optical responses is using a normalized L-norm loss against the optimal response $\y^*$, i.e., $\vert\vert \y - \y^*\vert\vert_l$ \cite{bezick2024pearsanmachinelearningmethod}. In simpler cases, when the response has to be maximized to some finite value- such as reflection,transmission coefficients or normalized power efficiency- the FOM can be directly defined as $\fom(\y) \equiv \y$. Deterministic formulations are the most common, since any finite sampling can be represented with a sufficiently larger vector $\y$.

Conversely, when the responses exhibit inherent variability due to noise, probabilistic metrics such as the KL divergence \cite{liu_hybrid_2020}, Reny-Divergence \cite{cai_multisensor_2019}, or Jenson-Shannon Divergence \cite{rodriguez-santos_identifying_2022} provide more robust alternatives by quantifying the statistical distance between observed and ideal optical response distributions, thereby capturing deviations that would otherwise not be apparent in deterministic methods.
Furthermore, neural networks have recently demonstrated alternative loss functions to better guide the optimization process, such as KID, FID, and perceptual loss \cite{bezick2024pearsanmachinelearningmethod, rombach_high-resolution_2022}.
While underutilized in PDD, these methods can effectively serve as learned FOMs, quantifying how closely a given response aligns with an optimal target. These networks are trained on large classification datasets and the loss is computed as an L-norm distance between activations in the deeper layers of the network, which encode important structural similarities. Neural network-based losses align well with human intuition for differences in data, however they can be much slower than the aforementioned losses.

\subsection{Latent Optimization}
The design space for PDD problems $\mathcal{X}$ is frequently exponentially large with a sparse distribution of useful designs; for example, the space of useful material height maps $\mathcal{X} \subset [0,1]^{n \times n}$ usually has rounded features, axial symmetries and minimum feature widths, all of which are sparsely distributed around $[0,1]^{n \times n}$.
In which case, directly optimizing the designs through adjoint methods is too costly and slow, as most variations are suboptimal with small, but computationally expensive step sizes.
Instead, designs can be compressed into lower-dimensional \textit{latent} feature vectors $z \in \mathcal{Z}$ using an encoder $q_\theta(z \vert \x)$ such that searching the associated latent space is more tractable using adjoint, gradient and global optimization techniques.
To use this more efficient design space, we train a decoder $p_\theta(\x\vert z)$ and modify our design objective in Eq. \ref{eq:des_opt_fom} to employ \textit{ latent optimization} methods  \cite{kudyshev_global_optimization, bezick2024pearsanmachinelearningmethod} to generate the optimal design 
\begin{align}
    \mathbb{E}_{z \sim q_\theta(z)}[\mathbb{E}_{x \sim p_\theta(x \vert z)}[\fom(\modelDetNum)]]. \label{eq:des_opt_latent_fom}
\end{align}
Here, the choice of optimizer $q_\theta(z)$ and decoder $p_\theta(x \vert z)$ are especially important, as they direct the performance.
While the design space $\mathcal{X}$ is typically sparse, the latent space $\mathcal{Z}$ is very dense, making it more efficient to explore.
For example, introducing continuous variations on the optical response $\hat{y}_\phi( x^{(0)} + \delta_x(t))$ by perturbing latent vectors $z^{(t)} \leftarrow z^{(0)} + \delta_z (t)$ allows the exploration of the design space $x^{(t)} \sim p_\theta(x^{(t)} \vert z^{(t)})$.
By adding subtle but impactful perturbations to latent representations of complex structures through global optimization \cite{wilson_machine_2021}, evolutionary and genetic algorithms \cite{liu_hybrid_2020}, differential evolution \cite{kudyshev_global_optimization}, and even ML-assisted optimization \cite{bezick2024pearsanmachinelearningmethod} can efficiently explore the latent space and offer alternative designs outside of human intuition \cite{wang_mechanistic_modeling, wilson2024nonnativequantumgenerativeoptimization, bezick2024pearsanmachinelearningmethod}.
Typically, the optimizer $q_\theta(z)$ is chosen such that it is efficient to sample and is coupled to the FOM by a latent surrogate model $E_\theta(z)$ that is trained to predict the FOM $\fom(\modelDetNum) : \x \sim p_\theta(\x \vert z)$. 
Recent work by the authors has demonstrated that simple L-norm energy-matching losses, e.g., $\vert \vert E_\theta(z) - \fom(\modelDetNum)\vert \vert_l$, are difficult to train and too restrictive for optimization. 
Naturally, the neighboring correlations between FOMs is more important to capture in the surrogate model than exactly matching the FOM. 
Therefore, by using a Pearson correlation loss instead of an L-norm, the optimizer can more efficiently sample $E_\theta(z)$.
The optimizer itself is often inspired by combinatorial optimization optimizers, such as, simulated annealing \cite{wilson_machine_2021}, global optimizers \cite{kudyshev_global_optimization}, recurrent neural networks \cite{bezick2024pearsanmachinelearningmethod} and even quantum samplers (Section \ref{sec:design:quantum}) \cite{wilson2024nonnativequantumgenerativeoptimization}.

\subsubsection{Variational Autoencoders (VAEs)} 

Variational Autoencoders (VAEs) are a type of unsupervised generative models that compress high-dimensional designs $x \in \mathbb{R}^n$ into low-dimensional latent vectors $z \in\mathcal{Z} \subseteq \mathbb{R}^d : d < n$ \cite{kingma_auto-encoding_2022}.
VAEs are often applied to denoising, compression, and optimization applications, where the feature rich latent vectors $z$ represent designs $x$ without its redundancies, noise, or symmetries which cannot be removed efficiently otherwise \cite{wang_disentangled_2024}.
The basic VAE workflow in ML-PDD begins with a set of optimized designs that is generated by classical methods, e.g., topology-optimized material distributions. Then, each design is compressed via an encoder $q_\theta(z \vert x)$ and eventually decompressed into the original design via a decoder $x \sim p_\theta(x \vert z)$.
The loss function is a two-term loss derived from the variational evidence lower bound
\begin{align}
    \mathcal{D}_{KL}(q_\theta(z\vert x) \vert\vert q(z)) - \mathbb{E}_{z \sim q_\theta(z \vert x)}[\log p_\theta(x \vert z)] \label{eq:VAE_Loss}
\end{align}
where $q(z)$ is a prior over $z$, $q_\theta(z \vert x)$ is the encoder, and $p_\theta(x \vert z)$ is the decoder. 
The first term $\mathcal{D}_{KL}$ is the Kullback-Leibler (KL) divergence which regularizes the distribution of the encoder with the prior $q(z)$ and the second term is the reconstruction loss that biases the generated design to be similar to the data design \cite{kingma_auto-encoding_2022}.

A significant advantage of using VAEs over diffusion models is their structured latent space $\mathcal{Z}$ given by optimizing the evidence lower bound in Eq. \ref{eq:VAE_Loss}.
When a VAE is properly trained \cite{dai_diagnosing_2019}, the decoder $p_\theta(x\vert z)$ efficiently constructs designs $x$ given latent vectors $z$ that are distributed according to the prior $q(z)$.
This dense distribution of latent vectors around the prior is what enables latent optimization to be so successful because small perturbations around the prior have a large impact on the design.

However, the trade-off between training the reconstruction loss and KL divergence terms often results in suboptimal generative performance, particularly when dealing with fine structural features.
Moreover, poor prior model assumptions $q(z)$, such as a single Gaussian distribution\cite{kingma_auto-encoding_2022}, can lead to poor regularization \cite{bezick2024pearsanmachinelearningmethod}, blurry reconstruction \cite{wilson2024nonnativequantumgenerativeoptimization} and difficulties in capturing multi-modal distributions for multi-objective problems.
For more advanced photonic design problems, such as nonlinear devices \cite{raju_frequency_doubling}, plasmonic devices \cite{masson_machine_2023}, or multi-objective devices \cite{kudyshev_lightsail_optimization}, these problems become more exaggerated and difficult to overcome. Several advanced strategies have been used to address these deficiencies in design.

First, conditional VAEs (cVAEs) incorporate conditional labels into the latent space as a latent feature, guiding the decoder to generate more specific properties on-demand, such as optical, structural, and material requirements, which must be incorporated into the inverse design process.
cVAEs are effective because the latent conditional label is also feature rich label and contains a lot of information for the decoder.
Secondly, cVAEs are well-suited for managing one-to-many mapping challenges by conditioning the model on desired outputs, enabling it to produce diverse solutions that meet the same optical response criteria.
This flexibility is valuable in exploring designs that meet multiple requirements, as cVAEs generate solutions within specific constraints.
Thirdly, the introduction of $\beta$-VAEs weighs the KL divergence with a scalar $\beta$ to balance the influence of the prior on training. 

\begin{figure*}[t!]
    \centering
    \includegraphics[width=\textwidth]{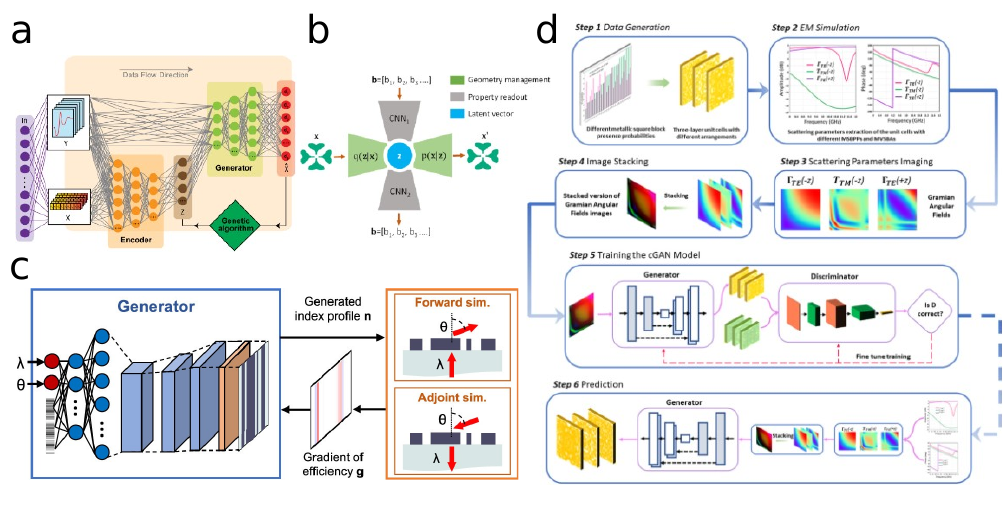} 
    \caption{\textbf{Variational Autoencoder (VAE) and Generative Adversarial Network (GAN) models in photonic design applications.} a) Architecture of GA-$\beta$-VAE for dielectric multilayer
structures inverse design. Adapted with permission from \cite{kumar_multi_solution}, Copyright 2024 Optica Publishing Group. b) A DNN Fusion model comprised of $\beta$-VAE and CNN1-z-CNN2 for inverse design and forward prediction, respectively, for bound states in the continuum (BICs) in freeform structures. Adapted with permission from \cite{li_bound_states}. Copyright 2021 Chinese Laser Press. c) Schematic of the conditional global topology optimization networks for metagrating generation. Adapted with permission from \cite{jiang_global_2019}. Copyright 2019 American Chemical Society. (d) The workflow of the cGAN-based methodology for the inverse design of multifunctional microwave metasurfaces. Adapted with permission from \cite{kiani_cGANs_metasurfaces}. Copyright 2022 The Authors. Advanced Photonics Research published by Wiley-VCH GmbH.}
    \label{fig:design}
\end{figure*}

For example, in Figure \ref{fig:design} (a), Kumar et al. implemented a constrained $\beta$-VAE model to optimize dielectric multilayer structures using genetic algorithms, allowing for multi-solution inverse design \cite{kumar_multi_solution}.
In another study by Lin et al., the authors examined eigenmodes for the inverse design of 2D BIC structures by using a $\beta$-VAE for encoding and decoding geometries, and two CNNs for forward simulation and inverse design (Figure \ref{fig:design} (b)) \cite{li_bound_states}.
By exploring the latent space of the $\beta$-VAE, the authors contributed to understanding protected/unprotected modes in complex geometries, enhancing future band engineering capabilities in photonic structures.

Moreover, using attention mechanisms allows the model to focus on small details over multiple passes \cite{vaswani_attention_2017, wilson24raptor}.
For instance, transformer-based architectures \cite{vaswani_attention_2017} leverage self-attention mechanisms to model long-range dependencies between data points, which is crucial for capturing complex interactions in photonic structures \cite{rane_transformers_2023}. 

Lastly, rather than relying on the KL divergence for regularizing the encoder, introducing an adversarial discriminator mechanism forces the latent space to better align with the prior through a discriminator model, thus improving the generative quality \cite{kudyshev_machine-learning-assisted_2020, kudyshev_global_optimization}.
Despite these improvements to VAEs, outside of latent optimization, they have largely been replaced by a more general framework which uses the aforementioned adversarial mechanism known as Generative Adversarial Networks or GANs \cite{zhou_gan_2023}.

\subsubsection{ Generative Adversarial Networks (GANs)}
\label{sec:design:gans}

Generative Adversarial Networks (GANs) attempt to address the regularization issues present in VAEs by eliminating the KL divergence loss entirely.
GANs consist of two main components, a generator $x \sim p_\theta(x)$ that produces designs, and a discriminator $d(x)$ that evaluates a design's ``authenticity'' $d(x) \in [0,1]$ with the training dataset \cite{goodfellow_generative_2020}, i.e., how probable is it that the data was constructed by the generator $d(x)=1$. Often the discriminator is trained but we omit its parameters. The adversarial training process iteratively refines the generator to produce increasingly realistic outputs to fool the discriminator.
On the other hand, the discriminator tries to accurately classify data to overcome the generator. So GANs' optimization objective can be expressed as a minimax problem:
\begin{align}
\min_\theta \max_d \mathbb{E}_{x \sim p_{\text{data}}(x)} [\log d(x)] + \mathbb{E}_{x \sim p_\theta(x)} [\log (1 - d(x))] \label{eq:GAN_Loss}
\end{align}

In contrast, VAEs minimize a combination of reconstruction loss and KL divergence. The KL divergence term regularizes the latent space, while GANs eliminate this term entirely, relying instead on the adversarial loss to guide the generation process. This difference in loss functions addresses the regularization issues in VAEs, but introduces challenges in balancing the generator and discriminator.

GANs are particularly valuable for developing device architectures \cite{liu_tackling_2021}, optimizing layouts \cite{qian_adaptive_2022}, enhancing imaging resolution \cite{you_ct_2020}, and solving inverse design problems \cite{christensen_predictive_2020,so_designing_2019,liu_intelligent_2021}.
They excel in searching large design spaces for configurations with desired optical properties, such as high-efficiency metasurfaces and low-loss waveguides \cite{jiang_simulator-based_2020,bui_design_2020} since they can learn complex mappings from design parameters to performance metrics and synthesize new designs that match the underlying data distributions, enabling efficient exploration of vast design spaces.
In inverse design, GANs offer a rapid alternative to traditional iterative methods.

However, GANs are prone to several challenges during training. Mode collapse occurs when the generator produces limited variations of outputs, failing to capture the diversity of the data distribution \cite{saxena_generative_2022}. Training instability refers to the difficulty in achieving a stable balance between the generator and discriminator, often leading to oscillating or divergent loss functions \cite{wiatrak_stabilizing_2020}.
To address challenges in training instability and mode collapse, advanced GAN variants such as Wasserstein GANs (WGANs) have been introduced. The Wasserstein distance measures the cost of transforming the distribution of generated data into the distribution of real data. Unlike the original GAN loss, which relies on binary cross-entropy, the Wasserstein distance provides a more meaningful gradient when the distributions of real and generated data are far apart \cite{deshpande_max-sliced_2019}. WGANs use this distance to improve the quality and diversity of generated designs, particularly in complex applications where structural nuances are critical \cite{gulrajani_improved_nodate}. For instance, Jiang et al. presented a global optimizer based on the WGAN model, as shown in Figure \ref{fig:design} (c), which can output ensembles of highly efficient topology-optimized metasurfaces operating across a range of parameters with efficiencies better than the best devices produced by adjoint-based topology optimization, while requiring less computational cost \cite{jiang_global_2019}.

Although GANs do not inherently impose structured control over the latent space, extensions such as conditional GANs (cGANs), similar to cVAEs, enable targeted design generation by incorporating specific constraints into the model, which are also adaptable to complex optimization challenges in photonics \cite{wang_high-resolution_2018}.
This capability allows GANs to address the one-to-many mapping problem, producing diverse designs that yield similar optical responses. 
For instance, Kiani et al. used a specialized cGAN \cite{kiani_cGANs_metasurfaces}, where the generator produces designs based on set conditions, such as specific optical properties (Figure \ref{fig:design} (d)).
They employed Gramian Angular Fields to transform optical property data into images, representing data correlations effectively.
This technique allows the creation of varied metasurface structures with similar electromagnetic functionality, handling the one-to-many problem well.

GANs also benefit from hybrid strategies that integrate their generative strengths with other models' latent space control by designing hybrid frameworks that combine adversarial loss for realism with reconstruction or noise-based loss for structured latent space and diverse sample generation. 
For instance, combining GANs with VAEs \cite{wang_generating_2022} or diffusion models \cite{wang_diffusion-gan_2023} enhances their ability to produce stable, high-quality designs. Diffusion models, in particular, provide noise-based generation processes that complement GANs, enabling diverse and realistic designs [11]. This integration could be particularly effective for photonic applications requiring high fidelity and adaptability in the generated outputs.

\subsubsection{Diffusion Models} 

While GANs are able to generate high-quality results, they still suffer from training instability due to their sensitivity towards hyperparameters and the balance of the relative power of the generator and discriminator.
Furthermore, VAEs are known to have issues balancing terms in the evidence lower bounds and their generative performance is lacking.
A class of models which recently gained popularity are denoising diffusion models.
Diffusion models are far more stable to train and robust towards hyperparameter changes as compared to GANs, and they have been shown to possess great generative capabilities and superior distribution coverage \cite{dhariwal_diffusion_2021}, outperforming GANs as evaluated by generative metrics such as Fr\'echet inception distance (FID).
Diffusion models have found applications in various domains, from computer vision to bioinformatics \cite{yang_diffusion_2023, cao_survey_2024}.
However, diffusion models are often more computationally expensive during training and sampling as compared to GANs and VAEs, requiring many more training steps to generate an image \cite{dhariwal_diffusion_2021}.

\begin{figure*}
    \centering
    \includegraphics[width=0.8\textwidth]{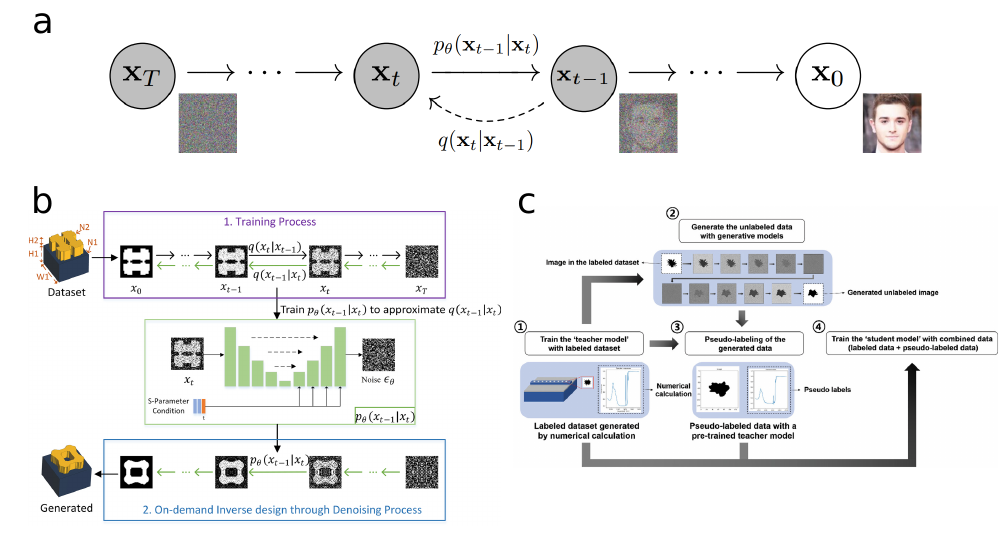}
    \caption{\textbf {Diffusion model applications in photonic design.}
    a) Illustration of reverse process, where the neural network learns how to denoise across the Markov chain to generate novel images. Adapted with permission from \cite{ho_denoising_2020}, Copyright 2020 Neural Information Processing Systems Foundation, Inc.
    b) Training architecture of conditional diffusion model, where a conditional and unconditional U-Net are mixed to produce new inverse designs through the denoising process. Adapted from \cite{zhang_diffusion_2023}. Copyright 2023 the author(s), published by De Gruyter, Berlin/Boston, licensed under the Creative Commons Attribution 4.0 International License.
    c) Illustration of semi-supervised training process. Adapted from \cite{kim_semi-supervised_2024}. Copyright 2024 the author(s). Laser \& Photonics Reviews published by Wiley-VCH GmbH, licensed under a Creative Commons Attribution-Non Commercial License.
    }
    \label{fig:diffusion}
\end{figure*}

A diffusion model uses a Markov chain $q(x^{(t+1)} \vert x^{(t)})$ to gradually add noise, e.g., Gaussian noise $x^{(t)} + \epsilon(t) : \epsilon(t) \sim \mathcal{N}(\mu_t, \sigma_t)$, to the training data $x^{(0)} \sim p_{\text{data}}(x)$ \cite{sohl-dickstein_deep_2015} so that the designs converge to a noisy Gaussian over time \cite{chan_tutorial_2024, ho_denoising_2020}.
A neural network predicts previous image $q_\theta(x^{(t-1)} \vert x^{(t)})$ to reverse the noising process, which during training iteratively predicts the previous image given the current image over all time steps.
To sample from the diffusion model, isotropic Gaussian noise is generated at the final time step $T$ and the neural network iteratively generates the previous image for $T-1$ time steps until a novel image is produced \cite{ho_denoising_2020, luo_understanding_2022}. 
To improve on the iterative denoising process, Ho et al. \cite{ho_denoising_2020} introduced the Denoising Diffusion Probabilistic Model (DDPM), where they parameterize the neural network to predict the original source of noise $\epsilon_\theta(t)$ instead of the previous image, which is shown in Figure \ref{fig:diffusion} (a). 
Typically for image applications, DDPMs use U-Nets which are deep convolutional networks with a spatial dimension bottleneck and skip connections to preserve information \cite{ho_denoising_2020, chen_overview_2024, ronneberger_u-net_2015}. 
Recently, latent diffusion models have been introduced, which perform the diffusion process within the latent space $\mathcal{Z}$ of a pre-trained autoencoder, reducing the training and sampling computational requirements as the expensive diffusion process occurs on a lower dimensional representation \cite{rombach_high-resolution_2022}.

Diffusion models have been applied to a variety of tasks in nanophotonics, such as metasurface design, classification, and image deconvolution \cite{zhang_diffusion_2023,kim_semi-supervised_2024,chakravarthula_thin_2023,zhu_meta_atoms,sun_photonic_2024}.
Zhang et al. \cite{zhang_diffusion_2023} focused on conditionally generating dielectric metasurfaces using a conditional diffusion model on an all-dieletric freeform unit cell dataset to generate high-quality designs, guided by S-parameters.
The researchers use a neural network to predict the spectral response and find that their diffusion model outperforms SLMGAN, WGAN-GP, and cVAE on frequency response accuracy, achieving a 43\%, 48\%, and 54\% improvement respectively. Their framework is shown in Figure \ref{fig:diffusion} (b). 
Kim et al. \cite{kim_semi-supervised_2024} sought to improve upon a classification task through implementing a semi-supervised learning strategy, facilitated by a diffusion model, that takes advantage of a large amount of unlabeled data with a small amount of labeled data, comparing their approach to a supervised learning scheme on a photonic crystal waveguide dataset.
Their strategy, as shown in Figure \ref{fig:diffusion} (c), first consists of training a ''teacher model'' to predict the spectral response of the generated dataset, and then training a DDPM to expand the original dataset, to which the "teacher model" affixes labels. Finally, a "student model" with the same architecture as the teacher model is trained on this expanded dataset. The study finds that this semi-supervised learning strategy can enhance average training losses of the student classification model by up to 102.8\% as compared to the teacher model.
Naturally, DDPMs are built for denoising. So, using them outside of design is also worthwhile.
Chakravarthula et al. \cite{chakravarthula_thin_2023} used a diffusion model to improve the optical quality of images captured from an on-sensor metalens array, a challenging problem prevalent at smaller wavelengths.
The authors introduce a flat array of metalenses to increase the field of view and implement a diffusion model to recover an image from the on-sensor metalens array. 
The image recovery process takes a captured image and alternates between inverse filtering, diffusion sampling steps, and merging steps.
Their study finds that their probabilistic deconvolution method outperforms existing traditional and more recent machine learning deconvolution methods in various metrics.

Diffusion models also provide an alternative approach to tackle one-to-many mappings in transforming random noise into structured designs. Zhu et al. demonstrated this method for meta-atom design \cite{zhu_meta_atoms}, using gradual noise-based generation specific to diffusion models to create multiple designs meeting the same specifications by varying the initial noise conditions. This allows the model to produce diverse solutions for the same performance criteria, which is ideal for cases where similar spectral responses arise from different structures. 

In the most recent studies, diffusion model-based multi-modal machine learning frameworks bridge the fields of natural language processing and computer vision by combining models that deal with these two domains. Sun et al. created a multi-modal machine learning framework that incorporated the Contrastive Language-Image Pre-training (CLIP) and Stable Diffusion (SD) models to predict the photonic modes in one certain structure from structural information text \cite{sun_photonic_2024}. The CLIP model is used to evaluate how similar a word is to an image, and the SD model performs the same diffusion process, but in the latent space of pre-trained autoencoders. The authors report that their method displays signs of stabilization across aHash, FID, and CLIP score metrics after only 6 training epochs. Furthermore, their method achieves an approximately 2-fold speedup in efficiency over conventional techniques that utilize Maxwell's equations.

\subsection{Reinforcement Learning}

Reinforcement Learning (RL) algorithms excel when design choices build on one another, making them well-suited for photonic design problems where each step can, in turn, drastically alter the optical response. They recently gained popularity for their ability to beat humans at chess, Go, and other popular games \cite{silver2018chessgo}. RL stands apart from other ML techniques because it uses a goal-driven, trial-and-error approach, where the RL agent learns by interacting with an environment and receiving feedback (rewards or penalties) based on its actions \cite{sutton2018reinforcement}. RL agents generate their datasets in real time instead of relying on large datasets at the outset of the training process. This reduces the number of simulations needed to meet specific performance benchmarks. Their autonomy and adaptability make RL uniquely suited to streamline photonic device design while minimizing the computational overhead required for training datasets.

\subsubsection{Reinforcement Learning Applied to PDD}
\begin{figure*}
    \centering
    \includegraphics[width=\linewidth]{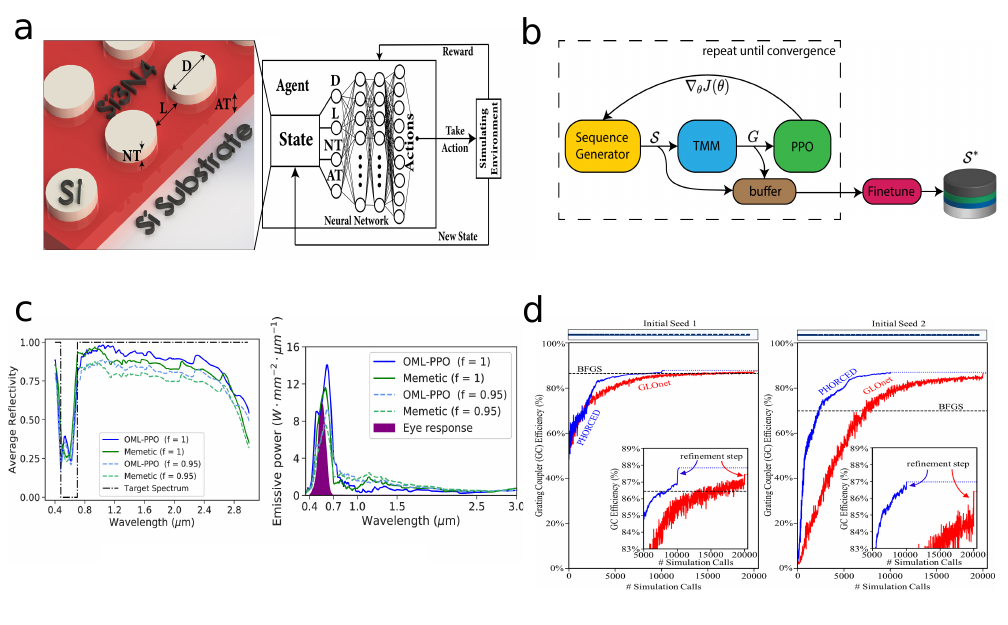}
    \caption{\textbf {Reinforcement learning for photonic design.} a) Example of reinforcement learning (RL) agent applied to the optimization of color generation using dielectric metasurfaces. Adapted with permission from \cite{sajedian2019optimisation}. Copyright 2019 Optical Society of America under the terms of the OSA Open Access Publishing Agreement. b) Training process for RL algorithm for the design of 1-dimensional photonic crystals. The Agent uses the Transfer Matrix Method (TMM) to simulate the spectrum of the design created by the sequence generator. Adapted with permission from \cite{wang2021automated}. Copyright 2021 The Author(s). Published by IOP Publishing Ltd. c) Result of the RL algorithm implemented in (b). The system outperforms the state-of-the-art memetic algorithm for the emissive power of light bulbs in the visible regime. Adapted with permission from \cite{wang2021automated}. Copyright 2021 The Author(s). Published by IOP Publishing Ltd. d) Silicon on Insulator (SOI) diffraction grating structure designed with RL algorithm. The RL model (PHORCED) outperforms GLOnet a compatible discriminative neural network trained with supervised learning. Both graphs in this subfigure adapted with permission from \cite{hooten2021inverse}. Copyright 2021 The Author(s). Published by De Gruyter, Berlin/Boston.
    }
    \label{fig:rl}
\end{figure*}

RL enables inverse design by forcing the agent to design a device environment $x$ (comprised of different materials and geometries) that optimizes the optical response, for example, the transmission spectrum of a metasurface \cite{sajedian2019doubledeep,so2020deep}. Figure \ref{fig:rl} (a) shows an example in which the device parameters $\x = \{\text{NT, L, AT, D}\}$ are the dimensions of the Si unit cells on a Si3N4 substrate \cite{sajedian2019optimisation}.
The agent iteratively adjusts these parameters to achieve desired optical response using feedback from a simulation $\hat{r}(\y \vert x)$ to generate positive and negative rewards. The agents' actions adjust the geometry or materials in the design environment $x$ and, over time, the RL model converges on optimal configurations. We simplify the RL training process to the following steps:

\begin{enumerate}
    \item \textbf{Initialize environment:} Create the input parameters $x^{(0)}$ for the policy function $\pi(x^{(0)})$. For example, in Figure \ref{fig:rl} (a), this means initially assigning values to the dimensions in the unit cell. The agent could also select from a list of materials although it is not included in this example.
    \item \textbf{Action Selection:} The agent uses its policy $\pi(x^{(t)})$ to determine the next action $a_{t+1} \sim \pi(x^{(t)})$, choosing from a range of probabilistically-generated actions based on the current state $x^{(t)}$. The agent balances exploration (trying new actions) and exploitation (repeating effective actions) as it refines the design.
    \item \textbf{Simulate the environment:} Each action’s outcome $\x^{(t+1)} \sim a_{t+1}(x^{(t)})$ is evaluated by simulating the optical properties $\hat{y}_\phi(x^{(t+1)})$, see Section \ref{sec:sims}. This simulation provides feedback on how well each action advances the design toward the target outcome.
    \item \textbf{Process Rewards:} The RL model estimates rewards (or penalties) for each action based on how closely the resulting design meets the target outcome. Then, these rewards (or penalties) are used to update the agent’s policy, in preparation for the next step.
    \item \textbf{Repeat:} Let the FOM $\fom(\y^{(t)})$ rank each design $\x^{(t)}$ by its optical response $\y^{(t)} = \hat{y}_\phi(\x^{(t)})$ from each generated design $\x^{(t)}$, for example using a loss function designed to minimize deviation from a target spectrum \cite{wang2021automated}. Iterate until the FOM $\fom(\y^{(t)})$ reaches a target threshold that signifies the otimization is complete.
\end{enumerate}

The previous architectures discussed in this review update their parameters based on some form of gradient descent incorporated with the results of a loss function, such as Eq. \ref{eq:VAE_Loss}. 
Similarly, RL techniques must specify a method for updating their agent’s policy; however, unlike the previously discussed architectures, this update is not necessarily derived from a loss function. Instead, RL techniques rely on feedback in the form of rewards (or penalties) to adjust the policy, based on the agent's progress toward achieving a specified goal.
This distinction has a significant implication: RL does not require the reward calculation process to be differentiable which means they can incorporate external simulations, such as 3D field solvers, into the optimization process.
This flexibility expands the range of problems that RL can address and allows researchers to integrate diverse computational tools into the training process

\subsubsection{Reinforcement Learning Architectures}
The two main policy update techniques in RL are called Q-learning, which is a value-based approach, and Policy Gradients (PG), which is a policy-based technique. Both methods offer distinct benefits for inverse design. 

The Q-learning method, first popularized in the Deep Q Network (DQN) model, updates an action value function $a_t(x)$ (or Q-values) for each state-action pair and evaluates them based on future rewards \cite{mnih2013playing}. The updates are processed using the Bellman equation in order to gradually improve Q-values by comparison of future rewards. The key to this technique is that it attempts to connect the state of the structure to the action that should be taken, making Q-values ideal for environments that have discrete parameter spaces.

Sajedian et al. use this process to find the optimal materials for maximizing the efficiency of a metasurface hologram \cite{sajedian2019optimisation}. They architect Q-values which map actions to the change of the material creating a state-action pair. They demonstrated this algorithm on a three-layer structure that was pre-designed to generate a hologram and successfully increased the efficiency of that hologram by 17\% using their RL algorithm. In another example, Badloe et al. design an ultra-broadband perfect absorber based on moth-eye structures. They expand beyond a discrete parameter space by using a Double Deep Q Network (DDQN) architecture that considers the height, width, and period of each nanostructure \cite{badloe2020biomimetic}. The key to this network is that it utilizes two networks that select actions and then evaluate them. Their approach demonstrates quick convergence using a range of different materials with over 90\% average absorption between 400 and 1600 nm for each design.

Alternatively, policy gradient methods (PMs) directly update the agent’s policy by adjusting its parameters to maximize rewards. Wang et al. demonstrate a PG technique for the design of one-dimensional photonic crystals \cite{wang2021automated}. They chose a Gated Recurrent Unit (GRU) architecture for their policy because it is a good representation of the sequential nature of multi-layer structures. Figure \ref{fig:rl} (b) shows a block diagram of their model. During each iteration, they use the GRU to select a series of materials for their structure, then they compute the result using the transfer matrix method (TMM), and then update the network using the Proximal Policy Optimization (PPO). PPO updates the policy by maximizing the expected reward, which, in this context, corresponds to how closely the generated optical spectrum matches the target spectrum. They designed an ultra-wide band absorber that increased efficiency from 95.37\% to 97.64\% between 400 to 2000 nm for a 5-layer system. They also designed an incandescent light bulb filter, using 42-layers, which achieved an enhancement factor of 16.60 demonstrating 8.5\% improvement over state-of-the-art techniques.

Several other papers highlight the versatility of policy gradient methods for the inverse design of photonic structures. Hooten et al. propose PHORCE (PHotonic Optimization using REINFORCE Criteria for Enhanced Design) which they use to increase the efficiency of a gradient coupler \cite{hooten2021inverse}. PHORCE is a great example of how RL can reduce the training time and training data required by other ML techniques. Figure \ref{fig:rl} (d) shows the performance improvement over a comparable deep neural network. Li et al. demonstrated a combination of Q-learning and policy gradient techniques for the inverse design of photonic crystals for nanoscale laser cavities \cite{li2023deep}. They found a crystal that achieved a Q-factor over 50 million, over 2 orders of magnitude better than state-of-the-art techniques.

RL has emerged as a powerful new technique for the inverse design of photonics structures. Unlike all other techniques discussed in this paper, it does not require a continuous function between the rewards it calculates and the updates to its policy. This frees up researchers to incorporate discontinuous analysis, like FDTD solvers or even real-world feedback into the training loop \cite{witt2023reinforcement,park2024sample}. Policy gradient methods are useful for continuous, high-dimensional action spaces whereas Q-learning excels at discrete action spaces, offering efficient solutions for problems with clearly defined actions. However, this freedom makes RL algorithms significantly more challenging to design, often requiring entire models to be built from scratch for a specific design. The resulting model is specific to a specific application and may not be as general as other ML techniques. They can also be computationally intensive, often requiring substantial training time and resources to achieve convergence over the other techniques \cite{witt2023reinforcement}. While RL presents computational challenges, its novelty in photonic design suggests significant potential for future applications as researchers continue to explore and refine its capabilities. 

\subsection{Quantum Generative Models} 
\label{sec:design:quantum}

\begin{figure*}[t!]
    \centering
    \includegraphics[width=\textwidth]{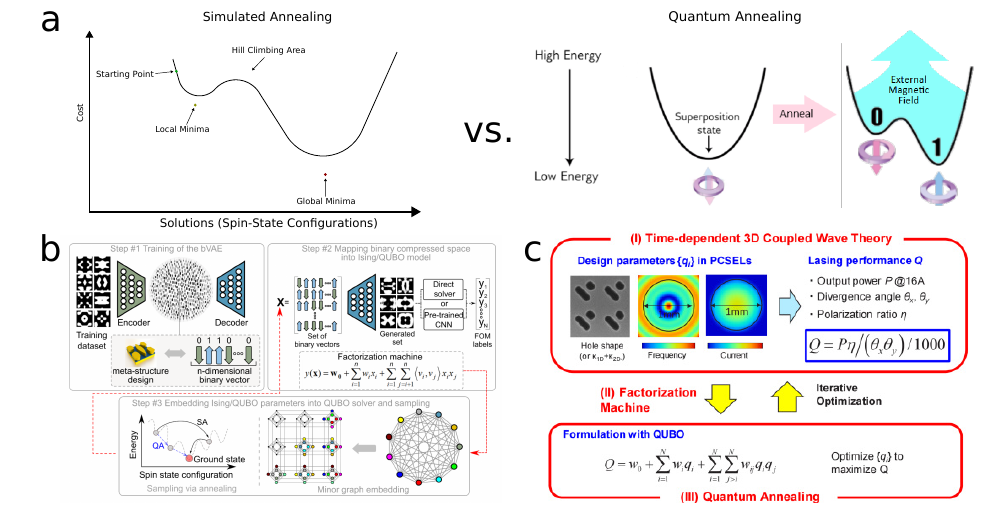} 
    \caption{\textbf{Quantum generative models in photonic design applications.} a) Simulated Annealing vs. Quantum Annealing. Simulated annealing aims to traverse a large solution space and converge at the global minimum as the solution. Quantum annealing works in a similar manner; it encodes the problem into a quantum system where the system's qubits represent binary variables. By evolving the system's Hamiltonian, all possible states of the qubits are explored simultaneously. Quantum annealing image adapted with permission from \cite{noauthor_algorithms_nodate}. Copyright D-Wave Systems. b) bVAE-QUBO Framework. Adapted with permission from \cite{wilson_machine_2021}, Copyright 2021 Author(s). Published under an exclusive license by AIP Publishing. c) Photon-crystal surface emitting lasers (PCSELs) using 3D RCWA, QUBO, and FMs. Adapted with permission from \cite{inoue_towards_2022}, Copyright 2022 Optica Publishing Group under the terms of the Optica Open Access Publishing Agreement.}
    \label{fig:quantum}
\end{figure*}

Quantum information science is a rapidly growing field, with recent breakthrough demonstrations accelerating enthusiasm for its potential \cite{acharya_quantum_2024, bravyi_high-threshold_2024}. Quantum-enhanced sampling and optimization algorithms are speculated to yield at least quadratic speed-up \cite{farhi2014quantumapproximateoptimizationalgorithm} for combinatorial optimization; thus, there is increasing interest to apply quantum algorithms for optimizing designs in PDD. 
The core idea is that designs $\x$ are embedded in quantum states $\vert x\rangle$ and sampled using a quantum device $q_\theta(x) = \vert\langle x\vert U_\theta\vert\psi_0\rangle\vert ^2$ where $U_\theta$ is a unitary acting on the initial state $\psi_0$. 
Quantum sampling devices can be accessed commercially through D-Wave \cite{king2024computationalsupremacyquantumsimulation}, IBM \cite{Qiskit2025}, Quantinuum \cite{Quantinuum2025}, to name a few.
To sample designs using a quantum device, the pure states $\vert \psi \rangle$ of the native Hamiltonian $\hat{H}$ of the device must encode the design $\x$.
Likewise, the FOM is often encoded onto the low-energy states of a surrogate Hamiltonian $\hat{H}$ such that the energy $\langle x \vert \hat{H} \vert x \rangle$ is anti-correlated with the FOM $\fom(\modelDetNum)$.
Then, the quantum device implements an algorithm, such as quantum-enhanced MCMC \cite{layden_quantum-enhanced_2023, wilson2024nonnativequantumgenerativeoptimization}, quantum annealing \cite{king2024computationalsupremacyquantumsimulation}, or QAOA \cite{farhi2014quantumapproximateoptimizationalgorithm}, to sample low-energy states of $\hat{H}$ to produce designs that optimize the FOM \cite{wilson2024nonnativequantumgenerativeoptimization, wilson_machine_2021} (Figure \ref{fig:quantum} (a)).
Initial work using this scheme in PDD problems was based on problems whose designs could be naturally encoded on pure states of the surrogate Hamiltonian.
For example, in Figure \ref{fig:quantum} (c) Inoue et al. optimized photonic-crystal surface emitting lasers \cite{inoue_towards_2022}.
The authors observe that non-uniform spatial distributions would be useful in accommodating for varying lattice constants or hole shapes in photonic crystal devices \cite{lima_thomes_investigation_2020}. 
These parameters are formulated into a Hamiltonian sampling problem using a factorization machine and sampled by D-Wave's quantum solvers. 
Similarly, Kitai et al. \cite{kitai_designing_2020} explored the design of metamaterials for radiative cooling applications using quantum sampling and a factorization machine. 
Properties such as compositional inhomogeneity can be formulated into minimization problem for the quantum annealer's Hamiltonian, which identified candidates with high FOMs.
The resulting structures were evaluated through rigorous coupled-wave analysis for their radiative properties, and the results were iteratively used to refine the factorization machine model.
Direct mapping of large problems increases the complexity of the system Hamiltonian, making it susceptible to noise and errors, while also exacerbating local minima trapping and significantly lengthening sampling times. 
In an effort to reduce this complexity, Ye et al. explored a hybrid quantum-classical strategy using Mixed-integer linear programming (MILP), which optimizes material layouts within a continuous domain using linear equations with continuous and discrete variables based on Generalized Benders' Decomposition (GBD) \cite{nocedal_numerical_2006, munoz_generalized_2011}. This approach optimizes goals such as structural compliance and heat transfer efficiency while managing constraints such as local displacement \cite{liang_topology_2019}. GBD decomposes the original PDD problem into a sequence of MILPs, and uses material layouts from previous iterations to determine each new iteration, leading to faster convergence compared to classic direct optimization methods. The material is represented as a binary variable, and the formulations are tested on D-Wave's systems, and the implementations using GBD (GBD Splitting-Direct and GBD Splitting-CQM) predictably took less time to converge, taking 234.12 and 111.97 seconds  respectively \cite{ye_quantum_2023}. Linear programming reduces iterations to reach optimal solutions, yielding sharper designs with fewer grey areas. Iterative refinements provide direct mappings, individual optimal solutions, faster convergence, and simpler implementation on quantum computers, extending to other continuous optimization problems.
Despite the efficiency gains from MILP, the need for robust feature extraction and dimensional reduction persists.

For other PDD problems, such as freeform material distribution optimization, the designs cannot be embedded directly into pure states via direct mapping because there are too few qubits, the design space is too sparse or unknown FOM mappings onto the native Hamiltonian.
Sparse design spaces in PDD especially complicate the optimization landscape, leading to inefficient use of quantum resources as many sampled quantum states may correspond to sub-optimal solutions.
This further motivates the use of autoencoders: more particularly, binary variational autoencoders are useful in representing continuous optimization problems by traversing a latent space \cite{team_keras_nodate}.
To address these challenges, discriminative models based on encoders compress designs onto pure states $\vert z \rangle$ that can be natively sampled \cite{wilson2024nonnativequantumgenerativeoptimization}.
These frameworks typically begin with a classical pre-processing step, where the PDD designs $x$ are encoded $q_\theta(z \vert x)$ onto a ``native'' representation $z$, sampled via the quantum device $q_\theta(z) = \vert\langle z\vert U_\theta\vert\psi_0\rangle\vert^2$, and then decoded $x \sim p_\theta(x \vert z)$ \cite{wilson2024nonnativequantumgenerativeoptimization, Wurtz2024}.
The bVAE-QUBO framework proposed by Wilson et al. aims to optimize dielectric, free-form diffractive meta-grating for beam steering. \cite{wilson_machine_2021}. As shown in Figure \ref{fig:quantum} (b), the framework begins by compressing input designs into a binary latent space with a corresponding polynomial (QUBO) formalism. Quantum annealing models optical behaviors as state transitions, solving them as combinatorial problems where optimal configurations represent solutions. Statistical mechanics models enhance this approach by encoding metamaterial features like refractive indices and conductivity levels. Using D-Wave's Leap hybrid models and simulated annealing, bVAE-QUBO achieves 96.5\% and 96.7\% respective efficiencies. The framework was also applied to optimize the efficiency of a thermal emitter for TPV cells, maximizing power generation and minimizing radiation losses. The bVAE-QUBO framework uses a reduced dimension space for faster processing and traversal, ensuring feature-rich encoding through one-to-one mapping and generating a probabilistic distribution of all possible topologies.
Notably, these techniques have been applied to freeform metamaterial unit-cell design for radiative cooling applications, thermophotovoltaic energy recapture \citep{wilson_machine_2021}, diffractive meta-gratings \citep{wilson_machine_2021}, and even improving future quantum devices \citep{wilson2024nonnativequantumgenerativeoptimization}.
Generally, these techniques have shown marked improvements beneficial in fields like thermophotovoltaics, incandescent light sources, biosensing, microbolometers, and drying furnaces \cite{kitai_designing_2020}.

\subsection{Summary and Outlook}
ML is transforming photonic design by enabling efficient exploration of complex, high-dimensional spaces to optimize device performance. Advanced generative models like VAEs, GANs, diffusion models, and RL streamline the design process by balancing innovation, precision, and practical constraints. These tools enable breakthroughs in photonic systems by facilitating rapid iteration, addressing multi-modal challenges, and uncovering new physical phenomena. Emerging quantum/hybrid quantum-classical approaches further extend these capabilities, driving innovation in advanced photonic applications with unprecedented efficiency and scalability.
\section{Fabrication}
\label{sec:fab}

Fabrication bridges the gap between simulation and experimental demonstration and presents a critical and unique challenge in the PDD.
It is a complex, often iterative and time-consuming process that involves balancing stochastic sources of error, including material inconsistencies, fabrication defects, and processing-induced variations \cite{lu2013nanophotonic, molesky2018inverse}.
This issue is exacerbated in the design of metamaterials for nanophotonics, where structures must achieve sub-wavelength precision while often extending over dimensions that span hundreds of wavelengths. Even minimal deviations, on the order of nanometers, can cause substantial degradation in device performance, impacting key FOM metrics such as efficiency, bandwidth, and functional robustness \cite{wiecha2021deep}.
Fortunately, the emergence of machine learning in nanophotonics enables precise, data-driven improvements to the complex, probabilistic nature of fabrication processes.
In PDD, fabrication is a stochastic process $\modelFab$ with parameters $\eta$ where $\chi$ is our fabricated design and $x$ is the ideal design.
Notably, unlike other probabilistic models such as $p_\theta$ where $\theta$ is variational and can be trained via machine learning methods, the fabrication process's dependency on its parameters $\eta$ can't be written down in a way that allows us to always optimize them, e.g., table alignments, machine settings, etc. 
An ideal fabrication process $\modelFab$ reduces the variation, i.e., the entropy, in the fabricated designs given an ideal design.
For the optimal fabrication process, would then be deterministic such that every design is fabricated perfectly every time.  
To address fabrication robustness and minimize design variability in PDD, we first discuss classical, non-ML approaches such as smoothing and vaccination. We then explore more advanced ML-driven methods, emphasizing discriminative models and reinforcement learning techniques.

\subsection{Improving Reliability with Stochastic Processes}

\begin{figure*}
    \centering
    \includegraphics[width=0.95\linewidth]{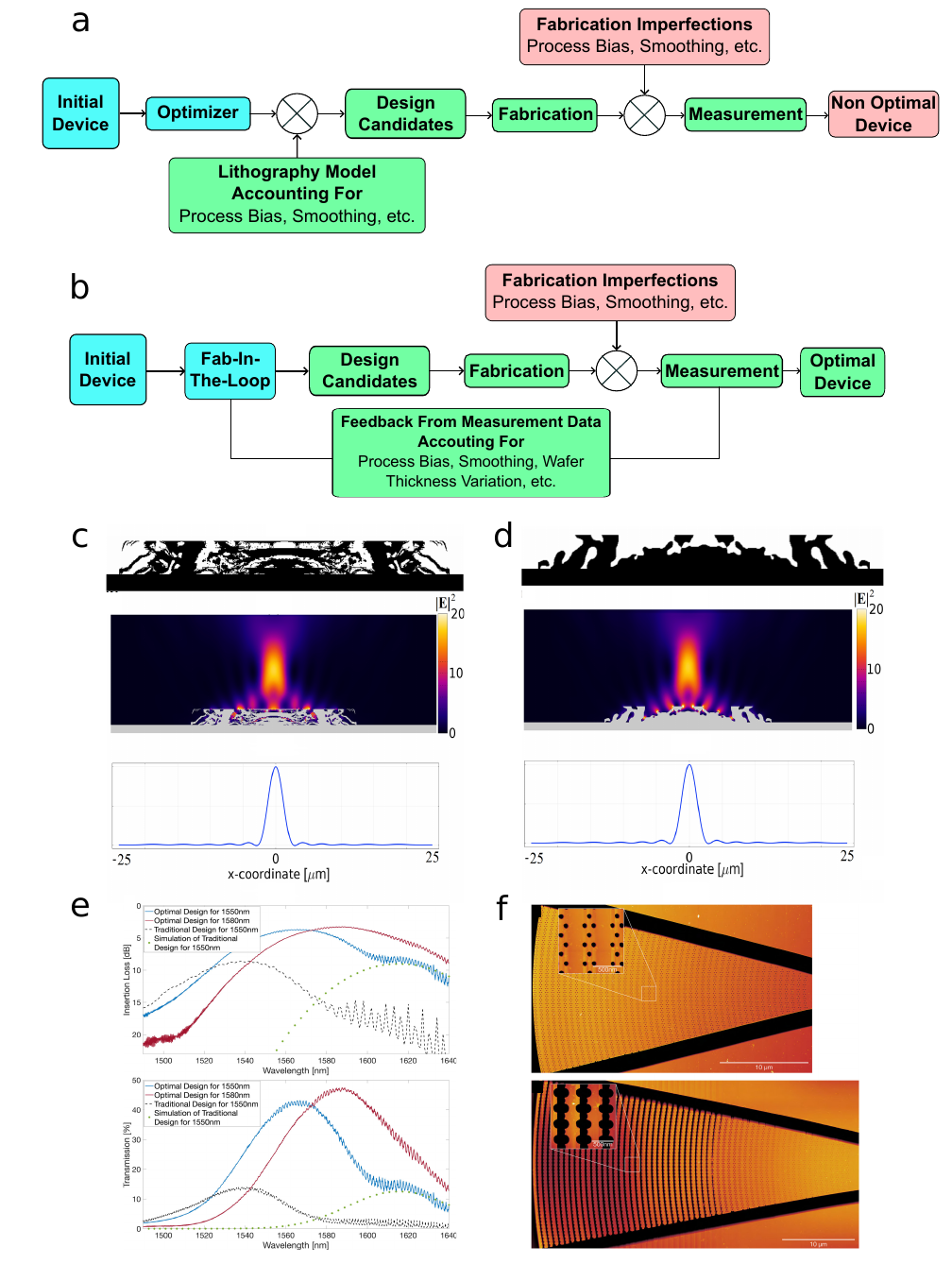}
    \caption{\textbf {Improving the fabrication of photonic devices with machine learning.} a) Inclusion of fabrication defects into the inverse design process. b) Reinforcement Learning (RL) uses a fab-in-the-loop process to optimize nanophotonic design over multiple iterations of the algorithm. c) Design of cylindrical silicon metalens using topological optimization. This example shows the result of the topological optimization without taking into account the fabrication process. d) The same design as (c) except that the topological optimization accounts for the fabrication process available to the designers of the system. Subfigures c and d are adapted with permission from \cite{Christiansen:21}, Copyright 2021 Optical Society of America. e) Result of the fab-in-the-loop process applied to a grating coupler design. f) Comparison between a traditional grating coupler and a grating coupler designed with the fab-in-the-loop RL algorithm. Subfigures a, b, e, and f are adapted with permission from \cite{witt2023reinforcement}, Copyright 2023 The Authors, distributed under the Creative Commons License, AIP Publishing.
    }
    \label{fig:fab}
\end{figure*}

Simple yet effective stochastic methods, such as smoothing, significantly enhance fabrication reliability. Smoothing typically employs Gaussian blurring ($\x \rightarrow \tilde{\x}$) on optimized structures generated by methods like topology optimization (Figure \ref{fig:fab}(a)) \cite{bendsoe2013topology, men2014robust}. This pre-fabrication step mitigates vulnerabilities to small-scale defects by reducing the complexity of fragile or unintuitive geometries frequently generated during optimization \cite{wang2011robust}. Figure \ref{fig:fab}(c) and (d) demonstrates the effect of smoothing on a cylindrical metalens designed via topological optimization, illustrating improved manufacturability and consistency between simulated and fabricated results. Smoothing techniques have been successfully applied to various applications, such as improving waveguide mode matching and enhancing robustness against defects through structural optimization \cite{khan2024fabrication}.

Vaccination is a similar technique that shifts the focus from the probabilistic incorporation of fabrication errors to adjusting misalignments common during experimental demonstration of free space optical designs, i.e., refining the fabrication parameters $\eta$ to reduce error. Mengu et al. demonstrated this technique on their Deep Diffractive Neural Network (D2NN) platform \cite{mengu2020misalignment}. Their inverse design process created a series of metasurfaces in which each unit cell was trained like a neuron in an artificial neural network. Each neuron's precise control and individual freedom left the platform vulnerable to misalignments on the order of a single wavelength. With the introduction of vaccination, the researchers introduced random misalignments in the training process, resulting in a robust experimental demonstration at up to four wavelengths. This technique is another promising example of how careful incorporation of noise during the training process can be generally helpful in the realization of nanophotonic devices \cite{montes2024fundamentals}.

\subsection{Correcting Variations with Discriminative Models}
While these simple stochastic processes are effective, researchers have sought more advanced techniques to protect against fabrication variations by adjusting the fabrication parameters $\paramFab$.
As we have demonstrated throughout this review, a natural next step to the modeling of complex stochastic processes is to use highly effective discriminative models which can model complex, nonlinear fabrication processes.
A common way to improve fabrication via discriminative modeling is to predict the optimal parameters $\paramFab^*(\x)$ to improve the fabrication process $\sModelFab_{\paramFab^*}(\chi \vert x)$ for any design $\x$.
For example, Gostimirovic et al. presented a feed-forward CNN to automatically correct layout distortions, which ensured that the resulting experimental demonstrations were robust against fabrication defects \cite{gostimirovic2022deep}. They trained their model on a small set of scanning electron microscopy images, enabling it to predict and then fix local variations such as corner rounding, bridging of narrow features, and over-etching in convex corners.
Their approach requires no modifications to the existing fabrication process or proprietary foundry data, making it well-suited to a wide variety of photonic applications. 

Alternatively, by predicting the error $\epsilon(\chi \vert \eta, x)$ in the fabricated designs $\chi$, adjustments can be made to the fabrication parameters $\eta$ to improve robustness.
In later work, the same group leveraged a deep CNN to predict nanoscale deviations from the original layout, revealing how proximity effects and other lithographic or etch-related artifacts degrade fidelity \cite{gostimirovic2023improving}.
In a similar approach, Liu et al. use a tandem architecture of deep neural networks to address disparities between their design simulations and optical response \cite{liu2018training}.
They address inverse design for problems like thin-film filters and multiwavelength phase modulators. They combine the simulation, design, and fabrication steps of the PDD into a single feedback loop using multiple machine learning approaches.
First, a discriminative model is trained to map a device layout to its optical behavior, then they combined the input design with a model that generates candidate layouts for a specified target performance. Although they set out to optimize their inverse design pipeline, their strategy naturally enhances robustness against manufacturing defects because the forward model is trained on fabrication data and the inverse network consistently generates layouts less prone to fabrication-induced errors. The resulting layouts avoid, for example, unrealistically thin layers that could cause over-etching or dimension loss. 

\subsection{Reinforcement Learning in Fabrication}

RL is a strong candidate to improve fabrication errors because of its naturally iterative and adaptive nature which real-time feedback from the fabrication process. This makes it well-suited to handle the complexities and uncertainties inherent in manufacturing at the nanoscale. One of the critical advantages of RL is its ability to model the fabrication process as part of the environment in which the learning agent operates. Witt et al. demonstrated a fab-in-the-loop RL algorithm that automates adjustments to the design based on measured results (Figure \ref{fig:fab} (b)) \cite{witt2023reinforcement}. The algorithm iteratively adjusted the design based on measured performance after each fabrication cycle. Specifically, the RL agent received feedback from insertion loss measurements and constructed rewards based on the results. After only six iterations of this process, they reduced insertion loss in a crystal grating coupler from 8.8 to 3.24 dB. In an alternative approach, Park et al. created a physics-informed RL model to design a reward system to encourage the feasibility of their fabricated devices \cite{park2024sample}. Figure \ref{fig:fab} (e) and (f) highlight the potential of RL for scalable, robust, and innovative solutions in the general design of nanophotonic systems. RL models provide a very general framework for machine learning, so introducing something like fabrication variations into the reward process can be as simple or complicated as the experimental demonstration may require.

\subsection{Summary and Outlook}
These fabrication-oriented methods showcase the growing capability of machine learning to address the randomness and complexity of real-world manufacturing conditions. Basic smoothing strategies provide fast, practical ways to refine the complex designs that sometimes result from inverse design techniques. Approaches like vaccination can ensure robustness against stochastic processes such as misalignment that might be present in experimental demonstration. Discriminative networks are an even more effective approach to protect against fabrication variations but they require datasets that can be time consuming to generate. Finally, RL provides a closed loop approach that can include fabrication results within the design process. Taken altogether, these techniques highlight a compelling trend. The inverse design of photonic structures push for tighter tolerances and higher complexity but machine learning techniques also offer a protective framework to ensure the manufacturability of those designs.

\section{Characterization}
\label{sec:characterization}
Once a design $\chi \sim \modelFab $ has been fabricated, its FOM $\fom(\y)$ is inferred $\hat{f}(\M)$ using a finite sample of noisy optical response measurements $\M = \{\m^{(i)} \sim \modelMeasure \}_{i=1}^M$, e.g., refractive index \cite{10.1063/5.0008026}, electronegativity \cite{Yanzhang23}, absorption \cite{PATEL2022112049}, chirality \cite{Kuznetsova24}, etc., from a measurement device $\modelMeasure$.
When measurement data can be rapidly collected, such as when simulations accurately mimic noisy measurements, or several measurements can be collected from a few fabricated devices, surrogate estimators can often significantly speed up characterization time, reducing the number of samples required to compute the FOM with a notable recent example being denoising back scattering in SEM \cite{KRISHNA2023113703}.
Moreover, AI-driven SEM \cite{Kandel23} in Figure \ref{fig:char} (c) combines quasi-random initial measurements with the Supervised Learning Approach for Dynamic Sampling using Deep Neural Networks to identify the optimal unmeasured points that, when added to the dataset, enhance the fidelity and quality of the reconstructed image.
Each unmeasured point is represented as a feature vector, with its elements determined by the measurement state in its surrounding neighborhood.
Additionally, there is substantial work being done in measuring thermal conductivity and corrosion degradation \cite{OJIH2023100286, XIA2022151, Nash2018, Coelho2022}, super-resolution imaging \cite{kudyshev2023machine}, authenticating the source of semiconductor devices \cite{wilson24raptor}, automating grain size measurements using GANs \cite{ANANTATAMUKALA2023113396} and detecting defects and cracks \cite{BramahHazela22, zhou23, zhou24, cranmer2023interpretablemachinelearningscience} using symbolic regression.
Known as short crack symbolic regression (SCSR) in Figure \ref{fig:char} (d), this knowledge-based method consists of three phases: data collection, domain knowledge-guided machine learning, and model extension.
Symbolic regression is involved by randomly generating the initial population and identifying individuals with high fitness to evolve through crossover and produce new offspring, as shown in the tree diagrams in Fig \ref{fig:char}. 
Notably, orders-of-magnitude speed-ups can be realized when characterization tasks require less accuracy, such as detecting single-photon emitters \cite{kudyshev2020sps} or tampering \cite{wilson24raptor}.
However, machine learning models are often limited by the quality and quantity of training data, the latter being a common problem for material science research and characterization \cite{horton21, KRISHNA2023113703, ZiatdinovHypo2022, wakabayashi22, Liu2023, Kandel23}.
In modern machine learning settings, there are often hundreds of thousands to millions of samples for training \cite{openai2024gpt4technicalreport}.
Recent advances in rapid fabrication techniques such as combinatorial spread libraries \cite{D4DD00109E, Potyrailo11}, human-automation workflows \cite{XIE2023101043}, and self-driving labs, enable larger data collection to improve characterization and experimental physical exploration. 
However, many of these methods are still in their infancy, and the more mature combinatorial spread libraries \cite{D4DD00109E} can only supplement this need with hundreds of samples a day when the measurements of interest are local, for example scanning probe microscopies.
In addition, the combinatorial search space is often exponentially large and intractable to search exactly \cite{ludwig19}.
To combat this low-sample problem, recent breakthroughs employ the following techniques which we explore further in this section:
i) generative modeling for pretraining
ii) active learning techniques using reinforcement learning and Bayesian optimization \cite{Liu2023, slautin2024measurementsnoisebayesianoptimization, 10.1063/5.0169961, Ghosh_2024, D3SC05281H}.




\subsection{Physics-Informed Generative Data Augmentation for Pre-Training}
\begin{figure*}
    \centering
    \includegraphics[width=\linewidth]{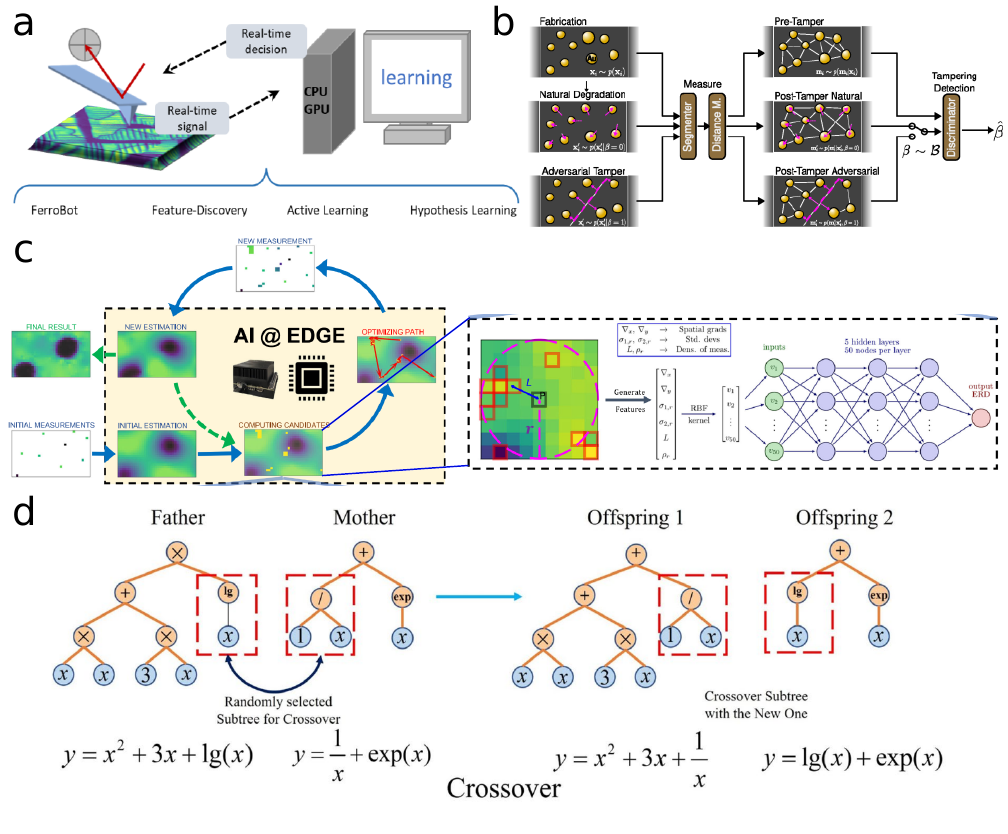}
    \caption{\textbf {Machine learning applications in photonic device characterization.} a) Automated Discovery using scanning probe microscopy. With the ubiquity of combinatorial spread libraries and scanning probe microscopy, active learning methods are required to acquire and interpret these spectroscopic measurements. The corresponding machine learning frameworks for automated SPM include FerroBOT and Feature-discovery, which follow predefined rules for decision-making. Reprinted with permission from \cite{ZiatdinovHypo2022}. Copyright 2022 American Chemical Society.
    b) Nanoparticle distance matrix characterization and discrimination using data augmentation and generative modeling. Adapted with permission from \cite{wilson24raptor}. Copyright 2024 SPIE Digital Library, Creative Commons Attribution 4.0 International License.
    c) AI Driven SEM. Adapted with permission from \cite{Kandel23}, Copyright 2023, UChicago Argonne, LLC, Operator of Argonne National Laboratory.
    d) Symbolic Regression in crack growth defects (SCSR). Adapted with permission from \cite{zhou24}. Copyright 2024, The Author(s) under exclusive license to The Korean Institute of Metals and Materials. 
    }
    \label{fig:char}
\end{figure*}

Two powerful techniques in machine learning for increasing training data are data augmentation \cite{mumuni2022data} and generative modeling \cite{bishop2006pattern}.
Typically, data augmentation increases the number of training samples by applying random transformations to the training data, e.g., additive Gaussian noise, image manipulation, etc.  
If the data augmentation is performed in a physics-informed manner, e.g., by applying generative modeling to learn the hyperparameters of noise sources during characterization, the generative models can be manually applied to pretrain a model on simulation data. 
For example, in a recent work by Wilson et al. \cite{wilson24raptor} as shown in Figure \ref{fig:char} (b), a set of 40 dark field microscopy images were augmented to generate a 10,000 image dataset by applying a pre-trained segmentation model to extract segmented images of gold nanoparticles spread uniformly on silicon. This model, along with a labeled clustering algorithm, are used to calculate the distance matrix and nanoparticle radii for adversarial tampering and natural degradation sample types. The discriminator network is then trained by randomly selecting a synthetic tampering type based on the tampering Bernoulli distribution.
The distribution of the nanoparticles was shown to be uniform, and the background Gaussian noise was measured. 
Taking the segmented nanoparticle images and measured background noise distributions, an arbitrary number of synthetic samples could mimic the characterization measurements by reproducing a uniform distribution of nanoparticles, each of which is randomly augmented with stretching, rotation, etc., and adding background noise. 
These methods work well when the physical model is well known, but the hyperparameters need to be learned.  
However, when the physical model is unknown and several candidates are available, we turn to alternative Bayesian methods.

\subsection{Experimental Physical Discovery}
The aim of automatic physical discovery \cite{Omidvar2024} is to predict an analytical model for a set of measurements \cite{ZiatdinovHypo2022, wakabayashi22, Liu2023} through Gaussian processes, active learning, and symbolic regression\cite{zhou23,zhou24,angelis23,cai2006heat, kim2023novel,li2022modeling, cranmer2023interpretablemachinelearningscience}.
Oftentimes, these models are employed when measurement data are sparse and time-consuming to collect and the physical model isn't known a priori.
For example, in hypothesis learning, a variant of active learning which uses reinforcement learning and Bayesian optimization, several candidate models are chosen with randomly initialized hyperparameters.
These models use a set of prior physical models $\hat{y}^{(i)}$ such as Gaussian priors and candidate models with preconfigured hyperparameters. 
Then, each of these models is evaluated using Bayesian inference based on reducing uncertainty.
Using a greedy policy, the model which has the minimal uncertainty is used to predict the next sample point $\phi^{(t+1)}$ and all the models are updated with the new characterization measurement.
Active learning enables algorithms to prioritize and identify critical structural or design features by iteratively refining their understanding based on user-defined parameters or feedback, such as key microstructural elements based on operator-defined signal aspects. Further, hypothesis learning optimizes functions using physical or competing models, such as optimizing microscope resolution or exploring ferroelectric domain growth \cite{ZiatdinovHypo2022} (Figure \ref{fig:char} (a)).
While hypothesis learning and other active learning methods have been demonstrated on small examples, as automated physical discovery matures, methods will need to be augmented and improved to address larger fabrication priors.
Along the same lines, active learning methods have few examples of integrating modern symbolic regression libraries for purely analytic modeling.

\subsection{Summary and Outlook}
Treating characterization as an inference task where the objective is to construct a symbolic model or approximate the FOM allows us to employ new techniques like symbolic regression and automated physical discovery.
To ensure that ML models are ready in low-shot environments like characterization, where sometimes a relatively low number of uniquely correlated samples are available, data augmentation techniques like generative modeling and random transformations can pre-train approximate FOM models.
The introduction of symbolic regression and automated physical discovery enable more efficient modeling of physical phenomena. 
The biggest future opportunities are in the introduction of community-wide nanophotonic databases that would give more characterization data for training these models.

\section{Conclusion and Future Outlook}

\subsection{Conclusion}
This work comprehensively reviews how machine learning-assisted photonic device development (ML-PDD) transforms the traditional, iterative photonic device development (PDD) approach. We broke down the PDD process into 5 steps: theory, simulation, design, fabrication, and characterization and showcase how machine learning (ML) algorithms improve each step of the process. This framework is presented in the context of discriminative models, which excel at representing sophisticated functions, and generative models, which excel at exploring large design spaces. Combining these techniques creates a data-driven approach that automates device exploration, speeds up the PDD process, and provides increasingly reliable real-world solutions.

The theory portion (Section \ref{sec:theory}) discusses how machine learning synergizes with the fundamental physics of electromagnetic theory to enrich the fidelity of photonic research. ML techniques like symbolic regression and discriminative models uncover and refine the underlying physical laws by proposing new, high-fidelity constitutive relationships in exotic media. Generative models provide insights into dominant mechanisms that aid performance through latent space engineering. Altogether, these developments pave the way for more comprehensive, accurate, and rapid electromagnetic modeling, linking photonic design directly to fundamental principles in ways that conventional analysis alone cannot achieve.

The simulation section (Section \ref{sec:sims}) highlights how ML approaches can assist or replace traditional simulation techniques like FDTD or RCWA. Discriminative models, which map design parameters to optical responses, act as fast, approximate simulators for specific problem areas. These techniques require extensive training data collected from numerical solvers or lab measurements. However, once trained, they can predict device behavior at a fraction of the time required by traditional techniques. Fortunately, research has shown that generative models can assist in data collection, offering data augmentation through synthetic examples that mimic actual or simulated measurements. This synergy helps to address the data scarcity problem that plagues many photonics workflows. Ultimately, the combination of these techniques is invaluable to the iterative optimization problems in photonic design, enabling the designer to search much larger design spaces than was previously possible while maintaining fidelity to established electromagnetic theory.

The design section (Section \ref{sec:design}) showcases the latest techniques for the inverse design of photonic structures using machine learning. We highlight the latest techniques for VAEs, GANs, diffusion models, reinforcement learning, and quantum-hybrid solvers. The generative models encode geometric features and material information into latent variables. Researchers can efficiently explore these expansive design spaces to refine design features and inversely optimize the solution. Reinforcement learning techniques provide a precise solution while minimizing the training data required to navigate the design space. Meanwhile, quantum-inspired and hybrid quantum-classical frameworks bring new frontiers to photonic optimization by leveraging quantum annealers or factorization machines for combinatorial challenges. All these ML-driven advances in design reflect a growing shift in the research community. Instead of manually guessing or tuning geometry, scientists are using neural networks to discover innovative device configurations that often outperform traditional techniques.

Next, we discuss how common imperfections in the fabrication process (Section \ref{sec:fab}), such as defects, misalignments, or tolerances, can be mitigated with modern ML techniques. We introduce straightforward solutions like smoothing techniques and “vaccination” against alignment errors. Then, we highlighted discriminative models capable of predicting lithographic distortions or etching errors, thereby proactively correcting design layouts for real-world production steps. Lastly, we introduce reinforcement learning techniques that include the fabrication process directly into the algorithm’s training environment, training the RL agent to propose robust designs against the fabrication process. We showcase algorithms at several levels of complexity, highlighting how the ML toolkit can be adapted to the fidelity required by the experimental demonstration.

The characterization section (Section \ref{sec:characterization}) emphasizes how ML can streamline data-heavy or noise-limited tasks. Discriminative models trained on partial information can quickly infer relevant properties when measuring the performance of complex photonic chips or metasurfaces. These models drastically reduce the need for exhaustive measurement protocols. These methods are critical for characterization, such as single-photon emitters, detecting subtle surface defects, or reconstructing near-field images. We then showcase how realistic noise models can ensure even more robust data augmentation techniques. Ultimately, ML-driven characterization enables more rapid insights into a device’s optical behavior, surpassing purely manual or physics-based interpolation.

Altogether, these new ML techniques demonstrate performance improvements across every step of the PDD. Discriminative models provide rapid forward mappings critical for speedy simulations, online process monitoring, and efficient characterization. Generative models offer unconstrained exploration of device topologies and parameter sets, which result in exotic yet robust design solutions. Combining both approaches in collaboration with physical knowledge and real fabrication constraints automates and accelerates the entire PDD process.

\subsection{Future Outlook}
In the short term, we expect that ML techniques will continue to improve the efficiency of photonics designs in terms of accuracy and cost. By incorporating physics-based models and fabrication constraints into the training process, they will ensure that designs are both physically sound and practically realizable. Hybrid modeling techniques that combine quantum annealers, latent space engineering, and high-fidelity surrogate models promise unprecedented efficiency in exploring complex design landscapes over today's solutions.

With regard to hybrid techniques that incorporate quantum annealers, however, it is evident that these tools are still in their incipient stage, and significant progress is required in both quantum hardware development and algorithmic design before their capabilities can be reliably leveraged in device design. In the foreseeable future, we anticipate the most viable role for quantum annealing to be a heuristic subroutine embedded within classical machine learning frameworks, offering possible advantages in constrained combinatorial tasks like QUBO, graph partitioning, or feature selection. Nevertheless, as Aaronson has emphasized, there are substantial hurdles like noise, limited qubit connectivity, and embedding inefficiencies that must be overcome to demonstrate the true utility of quantum optimization-assisted machine learning \cite{aaronson_read_2015}. Even so, incremental advantages in quantum materials, co-design, and surrogate modelling may gradually enable quantum optimization to become a more practical tool within machine learning.

As these approaches mature, we anticipate that the leading models will become more robust and expand to larger design spaces. Along this process, they will become increasingly accessible, making photonics designers less reliant on extensive machine learning backgrounds for selecting, training, and deploying models. Eventually, we expect these models to integrate directly with commercially available simulation software, similar to how topological optimization is already available in tools like Lumerical or COMSOL. We anticipate that generative inverse design frameworks will follow a similar trajectory toward maturity and eventually become a familiar and user-friendly tool employed across the majority of photonics designs.

Finally, we anticipate the creation and sharing of large datasets of photonic structures will standardize and expedite progress. We imagine a comprehensive, community-driven database similar the Materials project \cite{materialsproject13} and Atlas \cite{Hu2022}, which are already useful resources for the reference of optical data in homogeneous materials. Still, for the development of metamaterials \cite{metaset20} and similar complex photonic structures, there is a lack of a similar effort to construct a unified dataset of characterization measurements and computational models for training machine learning models. With leadership, the photonics community could compile the datasets created throughout their training processes. Then, as the field evolves and machine learning-assisted design becomes more prevalent, a central design database could vastly improve the PDD design cycle. Over time, ML will automate more and more of the PDD, improving technologies from integrated photonic circuits and topological photonics to advanced metasurfaces and quantum information systems.




\begin{acknowledgement}
Purdue team acknowledges the U.S. Department of Energy (DOE), Office of Science through the Quantum Science Center (QSC), a National Quantum Information Science Research Center, Air Force Office of Scientific Research (AFOSR) award No. FA9550-20-1-0124, Purdue’s Elmore ECE Emerging Frontiers Center ‘The Crossroads of Quantum and AI’, National Science Foundation (NSF) award DMR-2323910. The team at Northeastern University acknowledges the support from the NSF under Grant Nos. DMR-2202268, DMR-2323908, and ECCS-2430412. Georgia Tech team was supported in part by the NSF under Grant No. DMR-2323909, and in part by the Early-Stage Innovations (ESI) program of the National Aeronautics and Space Administration (NASA) under Grant No. 80NSSC23K0195 (subcontract from Baylor University; PI: Dr. Alan X. Wang). 
\end{acknowledgement}





\begin{conflictofinterest}
Authors state no conflict of interest.
\end{conflictofinterest}

\begin{dataavailabilitystatement}

\end{dataavailabilitystatement}

\bibliographystyle{ieeetr}
\newpage
\bibliography{refs}

\end{document}